# Analysis of Unsaturated Slope Stability under Seismic and Surcharge Loading by Upper Bound Rigid Block Method

Sumanta Roy[1,†,*]; Sourav Sarkar[2]; and Manash Chakraborty[2]

[1] *Dept. of Civil and Systems Engineering, Johns Hopkins University, Baltimore 21218, MD, United States.*

[2] *Dept. of Civil Engineering, Indian Institute of Technology (BHU), Varanasi 221005, UP, India.*

## Abstract

Failure of earthen slopes is a very recurrent phenomenon, credited mainly due to the excess rainfall and application of surfeit surcharge. However, most of the analyses regarding slope stability were performed without considering the unsaturated state of the soil. The prime purpose of the present manuscript is to address the stability of unsaturated homogeneous slopes subjected to surcharge load and pseudo-static seismic forces under different climatic conditions. The upper bound limit analysis technique was used based on the log-spiral failure mechanism. The suction stress-based effective stress approach was used to capture the effect of the unsaturated zone of the slope. The suction stress is modelled using Gardner's one-parameter hydraulic conductivity function and van-Genuchten's soil water characteristics curve. An extensive parametric study is carried out to assess the combined effect of slope geometry, soil-strength parameters, hydro-mechanical parameters, depth of water table, various flow conditions, surcharge load, and seismic loading. A few stability charts are proposed to show the impact of surcharge load and seismic load separately on unsaturated homogeneous slopes subjected to various climatic conditions. The present computed solutions match quite well with the available solutions prescribed in the literature.

## 1. Introduction

Although stability analysis of soil slopes has been studied for the last few decades, nevertheless, the problem still remains exciting and challenging on account of the various aspects of soil properties, climate change, innumerable complexities involved with the numerical simulations, and many other variabilities and uncertainties. Most of the previous analyses were accomplished by considering the extreme saturation state of the soil, namely, completely dry and fully saturated. However, in arid and semi-arid zones, the state of variable saturation state prevails up to a considerable depth below the ground surface. Therefore, unsaturated soil mechanics need to be employed to realistically account for the stability of soil slope. Moreover, the fluctuation of the water table and climatic conditions play a significant role in the soil properties of the vadose zone and gravely affect the overall stability of slopes. A good number of research in the past (Oloo et al., 1997; Costa et al., 2003; Vanapalli and Mohamed, 2007; Oh and Vanapalli, 2011, 2013) deliberately showed that the physiochemical and capillary forces developed in the vadose zone enhance the strength of the soil significantly. There are two-way approaches for analyzing the strength behaviour within the vadose zone: (a) matric suction based approach (Fredlund and Rahardjo, 1993; Travis et al., 2010; Gavin and Xue, 2010; Cho and Lee, 2001, 2002; Cai and Ugai, 2004; Hamdhan and Schweiger, 2013; Sun et al., 2016; Kang et al., 2020), and (b) suction-stress based approach (Griffiths and Lu, 2005; Lu and Godt, 2008; Vahedifard et al., 2015a; Vahedifard et al., 2015b; Vahedifard et al., 2016; Sun et al., 2019; Thota and Vahedifard, 2021). Considering the advantages of suction-stress over matric suction, the partially-saturated soil is modelled based on suction-stress-based effective stress equations. However, no literature seems to be available to assess the stability of unsaturated slopes subjected to surcharge and seismic loads. An effort has been exerted in this regard to evaluate the stability of homogeneous unsaturated soil slopes under surcharge and seismic loads by using the analytical upper bound rigid block method (UBRBM). The study is carried out for different flow conditions (hydrostatic, infiltration, and evaporation) and varying water table positions. An extensive parametric analysis was performed, and stability charts were presented for various slope geometry combinations, soil strength properties, hydro-mechanical parameters, water table depth, flow conditions, surcharge, and seismic loads.

---

[*] Corresponding author. Email: sroy41@jhu.edu
[†] The work was performed by the author while they were at Indian Institute of Technology (BHU).

## 2. Problem statement

Fig. 1 depicts a representative homogeneous and isotropic cohesive-frictional unsaturated soil slope on which the effects of seismic forces and surcharge pressures are verified. Following traditional and contemporary practices (Choudhury and Ahmad, 2007; Zhao et al., 2016), the seismic activity on the slope is idealized by the pseudo-static horizontal ($F_h=k_hW$) and vertical ($F_v=k_vW$) seismic forces; here, $k_h$ and $k_v$ represent horizontal and vertical seismic coefficients, respectively, and $W$ represents the weight of the failed soil block, as illustrated in Fig. 1a. The vertical seismic force is assumed to act in the downward direction. The form of the uniformly distributed surcharge pressure, $p_s$, chosen for the problem, is demonstrated in Fig. 1b. The geometry is represented by the (a) slope height $H$, (b) slope angle, $β$, and (c) ground surface inclination, $ω$. The groundwater table (GWT) is located at a depth of $h_w$ from the toe of the slope. The fluid flow is assumed to be unidirectional (vertical), under steady-state conditions, and governed by Darcy's linear flow and Gardner's (1958) hydraulic conductivity function. The variation of matric suction in the vadose soil zone is captured using the van-Genuchten (1980) SWCC model. The failure of the soil is dictated by the linear Modified Mohr-Coulomb (MMC) yield envelope as discussed in Roy and Chakraborty (2023). The MMC criterion features the additional strengthening effect of 'apparent cohesion' at every point in the vadose zone to some degree. For the implication of the limit theorem, the soil is further assumed to be perfectly plastic, and governed by associated flow rule. It is intended to evaluate the stability of the unsaturated soil slopes subjected to seismic pressure and surcharge loading by duly varying the hydro-mechanical properties, climatic conditions, and water table depth. The numerical simulations are carried out by writing suitable codes in *MATLAB 2018a*.

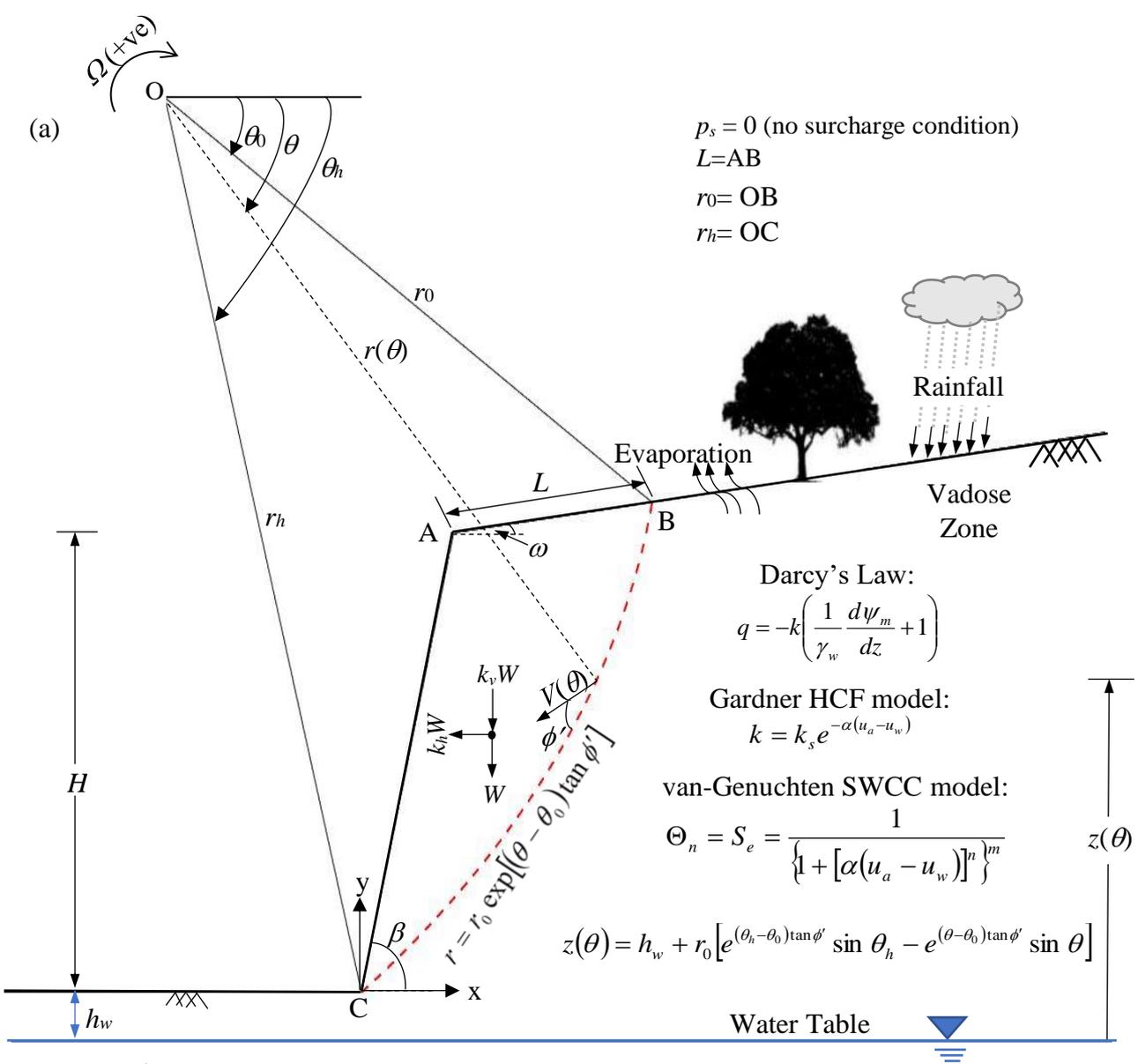

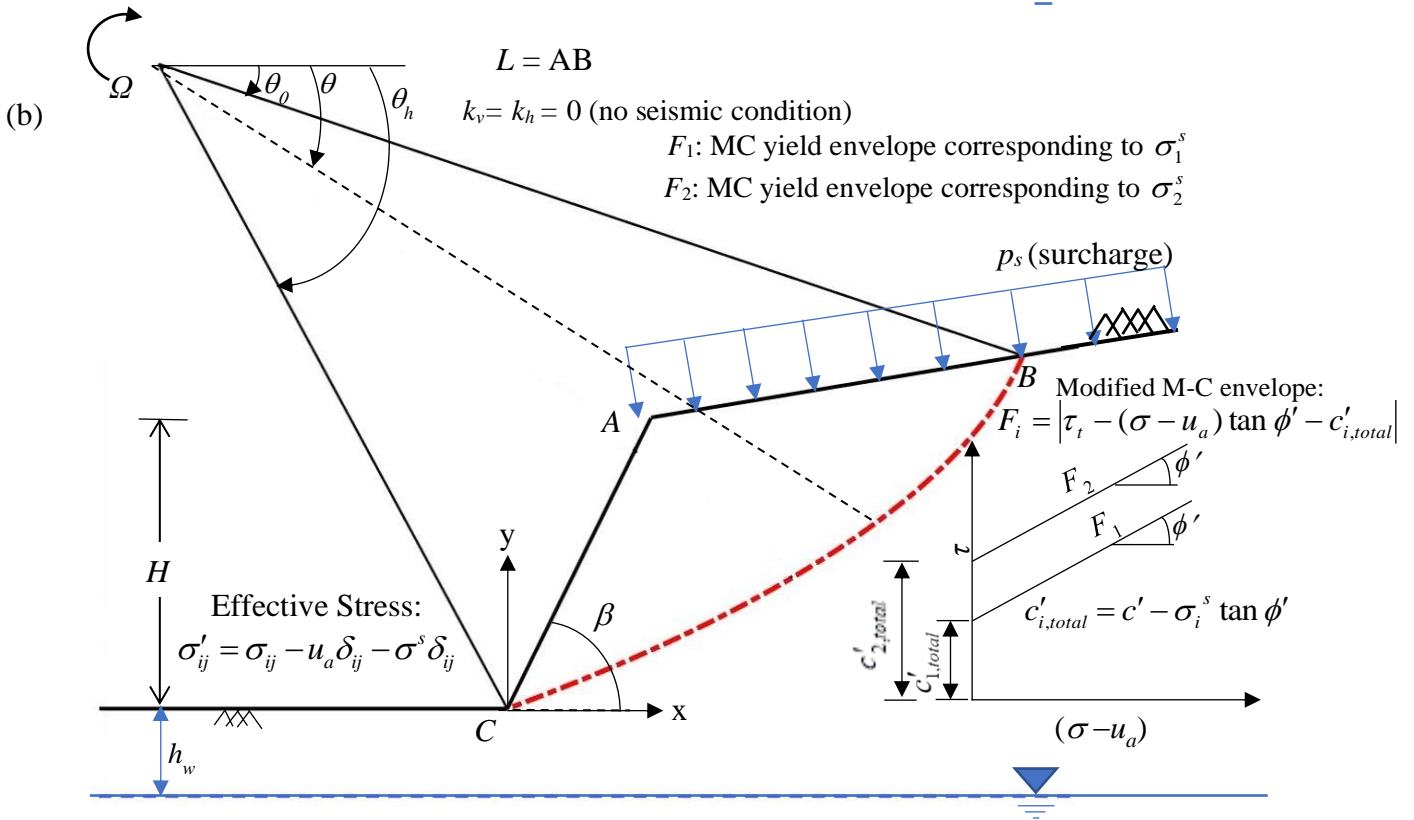

**Fig. 1.** Schematic representation of the chosen unsaturated soil slope of angle $\beta$ subjected to: (a) seismic activity without surcharge, (b) surcharge load without any seismic forces.

## 3. Modelling of unsaturated soils

The partially saturated soil is modeled by the following equations:

(a) van-Genuchten (vG) SWCC formulation: $\Theta_n = S_e = \left\{1+\left[\alpha(\psi_m)\right]^n\right\}^{-m}$ (1a)

(b) Darcy's (1856) linear flow law: $q = -k\dfrac{dh_m}{dz} = -k\left(\dfrac{1}{\gamma_w}\dfrac{d\psi_m}{dz}+1\right); h_m = \dfrac{\psi_m(z)}{\gamma_w}+z$ (1b)

(c) Gardner's (1958) HCF: $k = k_s e^{-\alpha\psi_m}$; $\psi_m(=u_a - u_w)$ is the matric suction (1c)

where, (i) $\Theta_n$ =normalized volumetric water content, $S_e$=effective degree of saturation; the value of $\Theta_n$ and $S_e$ ranges between 1 (i.e., completely saturated state) and 0 (i.e., residual state).

(ii) $u_w$ and $u_a$ are the pore water and air pressures, respectively; $\gamma_w$=unit weight of water; $k_s$= saturated hydraulic conductivity of soil.

(iii) $\alpha$, $n$, and $m$ are the vG-SWCC fitting parameters. $\alpha^{-1}$ indicates the air entry value (AEV), $n$ denotes the pore dimensions and their distributions, and $m$ indicates the overall symmetry of vG-SWCC curve. The value of $\alpha$ is reported to lie between 0.1-0.5, 0.01-0.5, and 0.001-0.01 for sands, silts, and clays, respectively (Lu and Griffiths, 2004); $n$ possess a value in the range of 1.1 to 8.5 (Singh, 1997; Lu and Likos, 2004), and the parameter is obtained through a strict relation ($m=1-1/n$), as demonstrated earlier by Mualem (1976) for obtaining a closed-form hydraulic conductivity function.

(iv) $z$ (positive upwards) represents any arbitrary distance above the water table; and

(v) *q* represents the flux conditions, which can be evaporative (positive), infiltrative (negative), and no flow (zero); the evaporation, infiltration, and no-flow conditions are abbreviated here as EV, IF, and NF respectively.

Integrating the combined form of Eqs. (1)a–(1)c and imposing the Dirichlet boundary condition (zero head at GWT) eventually results in the following expression of matric suction at any arbitrary position *z* above the water table:

$$\psi_m = -\alpha^{-1} \ln\left[(1+Q)\, e^{-\alpha \gamma_w z} - Q\right]; \quad Q \text{ (surface flow ratio)} = q/k_s \tag{2}$$

For the past few years, the suction stress, conceived by Lu and Likos (2004) is used as the fundamental stress state variable for several stability analyses. The advantages of using suction stress in the upper bound analysis are elaborately described in Prasad and Chakraborty (2023). Based on the thermodynamic principle, Lu et al. (2010) proposed the following closed-form suction stress equation:

Suction Stress: $\sigma^s = -\psi_m \Theta_n = -\psi_m S_e$ (3)

The merging of Eqs. (1)a, (2), and (3) yields the ensuing expression:

$$\sigma^s(z) = \frac{1}{\alpha} \frac{\ln\left[(1+Q)\, e^{-\alpha \gamma_w z} - Q\right]}{\left(1 + \left\{-\ln\left[(1+Q)\, e^{-\alpha \gamma_w z} - Q\right]\right\}^n\right)^{1-1/n}} \tag{4}$$

Eq. (4) represents the suction stress characteristic curve (SSCC) which indicates the spatial variation of the suction stress in the vadose zone. Fig. 2 demonstrates the variation of matric suction, effective saturation, and suction stress above GWT for two different AEV's. The suction stress profiles appear to be highly nonlinear for lower AEV soils. Following Lu and Griffith (2004) the value of *Q*, which accounts for infiltration rate, invariably remains less than unity, reflecting the assumptions of steady-state flow and numerical validity of the equations employed

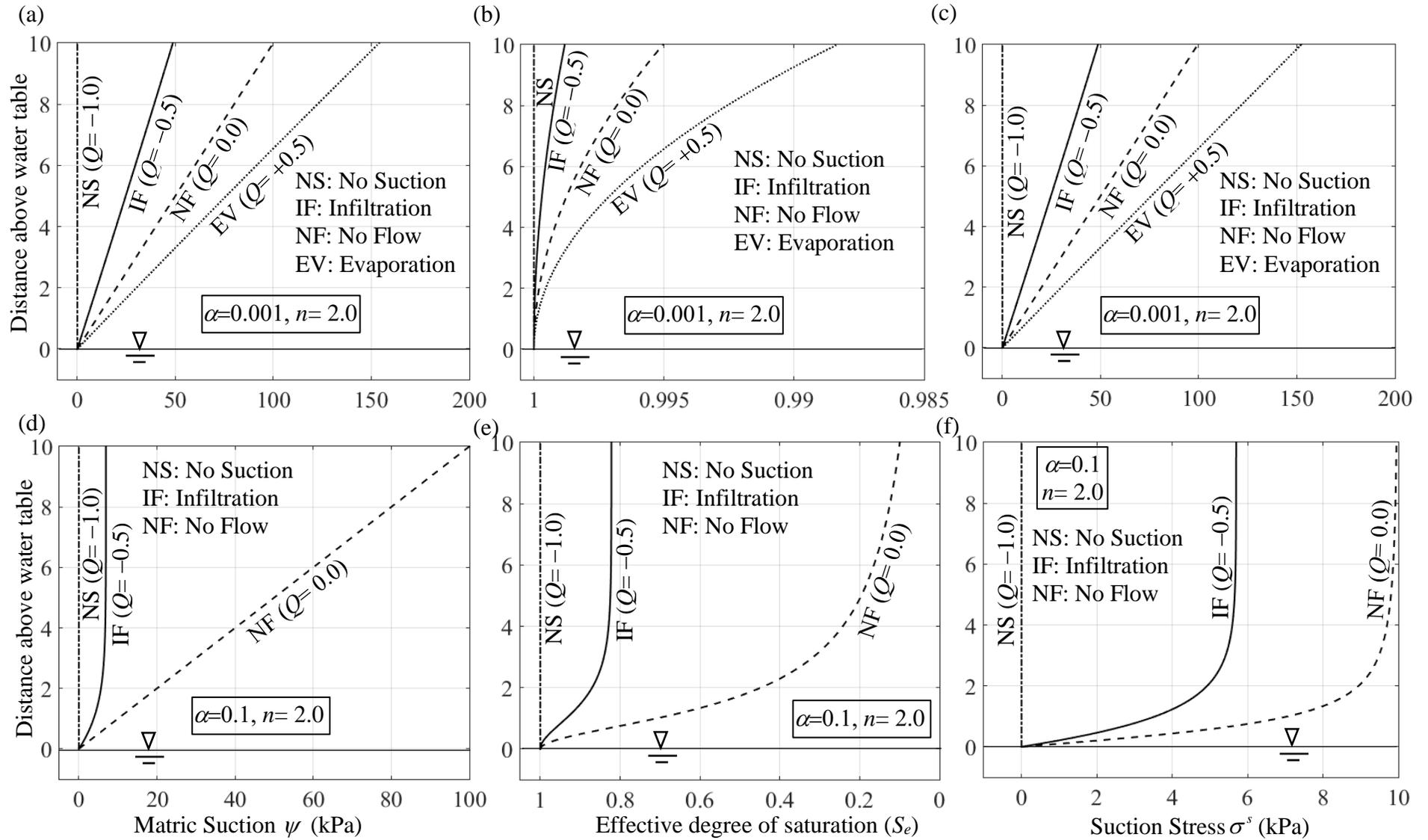

**Fig. 2.** The variation of matric suction, effective degree of saturation, and suction stress above GWT for two different air entry values: (a, b, c) $\alpha=0.001$ and (d, e, f) $\alpha=0.1$.

After the inclusion of the suction stress parameter, the MMC yield envelope can be represented as under:

$$\tau = c' + \sigma'_n \tan\phi' = c' + \left[\sigma - u_a - \sigma^s\right]\tan\phi' = c'_{total} + (\sigma - u_a)\tan\phi' \qquad (5)$$

Where, (i) $\sigma'_n$ (effective normal stress) $= \sigma - u_a - \sigma^s$ \hfill (6a)

(ii) $c'_{total} = c' + c'_{apparent}$ ; $c'_{apparent}$ (apparent cohesion) $= -\sigma^s \tan\phi'$ \hfill (6b)

Usage of the single stress state variable, as argued by Khallili and Khabbaz (1998), provides enormous flexibility in the computational processes. With the increase in suction stress, the MMC yield surface shifts parallelly upwards indicating an increase in the shear strength. The subsequent section elaborates the suction stress-induced MMC-based UBRBM formulations for obtaining the stability of the soil slopes by including the effect of pseudo-static seismic forces and surcharge loads.

## 3. Upper bound limit analysis formulation

### 3.1. Overview

The upper-bound theorem of limit analysis is a highly effective method of dealing with problems in geotechnical engineering (Chen, 1975; Sloan and Kleeman, 1995; Soubra, 1999; Zhu, 2000; Kumar and Samui, 2006; Kumar and Kouzer, 2008; Kumar and Chakraborty, 2014; Chakraborty and Kumar, 2015; Chen and Xiao, 2020). The rigid-block based upper bound stability analysis is performed by a suitable assumption of the kinematically admissible collapse mechanism. No such prior assumption of collapse mechanism is required in the conventional finite element–based limit analysis methods; nevertheless, the present technique is adopted due to its easiness, simplicity, and computational flexibility. The objective function is obtained by equating the rate of total external work ($\dot{W}$) with the total internal power dissipation ($\dot{D}$). This objective function is minimized to provide the upper-bound solutions and the geometrical configuration of the failure mechanism.

## 3.2 Collapse mechanism and formulations

Following the past studies (Vahedifard et al., 2016; Sun et al., 2019), the considered shallow slope is assumed to fail by developing a log-spiral toe failure surface that obeys the following equation: $r = r_0 e^{(\theta - \theta_0)\tan\phi'}$; here, $r$ is any arbitrary radius vector and $\theta$ is the subtended angle (measured clockwise) between the horizontal axis and $r$ vector. The sets ($r_0$, $\theta_0$) and ($r_h$, $\theta_h$) represent the first (OB) and last (OC) radius vectors and their corresponding subtended angles. Fig. 1 clearly describes the meaning of all the parameters for any arbitrary log-spiral mechanism having center point at O. The angular velocity rate of the sliding mass is indicated by $\Omega$ (clockwise considered positive). The log spiral meets the top ground surface of the slope at point B. The length of AB is denoted by $L$. By simple geometric relationship, the following two non-dimensional length ratios are obtained (Chen, 1975):

$$\frac{H}{r_0} = \frac{\sin\beta}{\sin(\beta - \omega)} \left\{ \sin(\theta_h + \omega) e^{(\theta_h - \theta_0)\tan\phi'} - \sin(\theta_0 + \omega) \right\} \qquad (7a)$$

$$\frac{L}{r_0} = \frac{\sin(\theta_h - \theta_0)}{\sin(\theta_h + \omega)} - \frac{\sin(\theta_h + \beta)}{\sin(\theta_h + \omega)\sin(\beta - \omega)}\left\{e^{(\theta_h - \theta_0)\tan\phi'} - \sin(\theta_h + \omega)\right\} \quad (7b)$$

For obtaining the upper bound solutions, the expressions of $\dot{W}$ require to be equated with the expressions of $\dot{D}$. The expressions of these terms are evaluated below.

### 3.2.1. Rate of external work done ($\dot{W}$)

In the presence of pseudo-static loading and the surcharge pressure, $\dot{W}$ constitutes the rate of work done due to the (a) weight of failed soil block ($\dot{W}_{weight}$), (b) pseudo-static seismic forces ($\dot{W}_{earthquake}$), and (c) surcharge loads ($\dot{W}_{surcharge}$).

The expression for each of the terms can be written as follows:

Considering Chen (1975) and Utili (2013): $\quad \dot{W}_{weight} = \gamma r_0^3 \Omega (f_1 - f_2 - f_3)$ \quad (8a)

Considering Zhao (2016): $\quad \dot{W}_{earthquake} = \gamma r_0^3 \Omega k_v (f_1 - f_2 - f_3) + \gamma r_0^3 \Omega k_h (f_4 - f_5 - f_6)$ \quad (8b)

By simple trigonometric relationship: $\quad \dot{W}_{surchage} = (p_s L)\Omega (r_0 \cos\theta_0 - 0.5L\cos\omega)$

$$= p_s \Omega r_0^2 \left(\frac{L}{r_0}\cos\theta_0 - 0.5\frac{L^2}{r_0^2}\cos\omega\right) \quad (8c)$$

Unlike $\dot{W}_{weight}$, the $\dot{W}_{earthquake}$ term involves the effect of the vertical as well as horizontal forces; nevertheless, the basic architecture of $\dot{W}_{weight}$ and $\dot{W}_{earthquake}$ remain the same.

### 3.2.2. Rate of energy dissipation

Due to the rotational sliding mechanism, the energy dissipated ($\dot{D}$) across the log spiral surface, CB, can be expressed as:

$$\dot{D} = \int_{\theta_0}^{\theta_h} c'_{total}(V\cos\varphi')\frac{rd\theta}{\cos\varphi'} = \int_{\theta_0}^{\theta_h}(c' + c'_{apparent})(V\cos\varphi')\frac{rd\theta}{\cos\varphi'} \quad ; V(\text{Linear velocity}) = r\Omega \quad (9a)$$

$$\dot{D} = \underbrace{\int_{\theta_0}^{\theta_h} c'(V\cos\phi')\frac{rd\theta}{\cos\phi'}}_{\text{Due to internal cohesion}\ (\dot{D}_{cohesion})} + \underbrace{\int_{\theta_0}^{\theta_h} c'_{apparent}(V\cos\phi')\frac{rd\theta}{\cos\phi'}}_{\text{Due to suction stress}\ (\dot{D}_{suction})} = \underbrace{\dot{D}_{cohesion} + \dot{D}_{suction}}_{\text{Dissipation Components}} \quad (9b)$$

Here, $V$ and $\Omega$ are the linear and angular velocities, respectively. The contributions to energy dissipations by the soil's internal cohesion and by the matric suction are separately analyzed and shown in Appendix A.2. The final decoupled form of $\dot{D}$ conforms to the following equation:

$$\dot{D} = \frac{c'r_0^2\Omega}{2\tan\phi'}\left[e^{2(\theta_h - \theta_0)\tan\phi'} - 1\right] + \int_{\theta_0}^{\theta_h}(-\sigma^s\tan\phi')\Omega r_0^2 e^{2(\theta - \theta_0)\tan\phi'}d\theta \quad (10)$$

The relationship between the vertical distance, $z$ and $\theta$ can be further expressed as:

$$z(\theta) = h_w + r_0\left[e^{(\theta_h - \theta_0)\tan\phi'}\sin\theta_h - e^{(\theta - \theta_0)\tan\phi'}\sin\theta\right] \quad (11)$$

After substituting the expression of $z(\theta)$ in the second term of $\dot{D}$ (i.e. $\dot{D}_{suction}$), the task of finding the closed-form integral solution becomes immensely complicated. Therefore, numerical integration is performed through Gauss Quadrature scheme.

### 3.3.3. Determining the upper bound stability number ($S_n$)

The stability number, defined as $S_n = \gamma H/c'$, (Chen, 1975), characterizes the overall stability of the slope; here $H$ represents the maximum slope height corresponding to unit factor of safety.

Applying the theorem of upper bound limit analysis, the stability number for the seismic and surcharge loadings are obtained below:

*For pseudo-static forces:* $\dot{W}_{weight} + \dot{W}_{earthquake} = \dot{D}$  (12a)

$$\Rightarrow \left.\frac{\gamma H}{c'}\right|_{ps} = \left( \frac{\sin\beta \left( \frac{1}{2\tan\varphi'}\left[e^{2(\theta_h-\theta_0)\tan\varphi'} - 1\right] + \int_{\theta_0}^{\theta_h}\left(-\sigma^s \tan\varphi'\right)e^{2(\theta-\theta_0)\tan\varphi'}d\theta \right)}{\left[\frac{\sin(\beta-\omega)}{\left\{\sin(\theta_h+\omega)e^{(\theta_h-\theta_0)\tan\varphi'} - \sin(\theta_0+\omega)\right\}}\right]\left[(1+k_v)(f_1 - f_2 - f_3) + k_h(f_4 - f_5 - f_6)\right]} \right)$$  (12b)

*For surcharge loads:* $\dot{W}_{weight} + \dot{W}_{surcharge} = \dot{D}$  (13a)

$$\Rightarrow \left.\frac{\gamma H}{c'}\right|_{sur} = \left( \frac{\frac{1}{2\tan\varphi'}\left[e^{2(\theta_h-\theta_0)\tan\varphi'} - 1\right] + \int_{\theta_0}^{\theta_h}\frac{1}{c'}\left(-\sigma^s \tan\varphi'\right)e^{2(\theta-\theta_0)\tan\varphi'}d\theta - \frac{P_s}{c'}\left(\frac{L}{r_0}\cos\theta_0 - 0.5\frac{L^2}{r_0^2}\cos\omega\right)}{\left[\frac{\sin(\beta-\omega)}{\sin\beta\left\{\sin(\theta_h+\omega)e^{(\theta_h-\theta_0)\tan\varphi'} - \sin(\theta_0+\omega)\right\}}\right](f_1 - f_2 - f_3)} \right)$$  (13b)

Eqs. (12)–(13) are highly non-linear. For the given slope profile and the hydromechanical parameters, $S_n$ depends on the collapse mechanisms defining parameters ($\theta_0$, $\theta_h$) as well as on $H$ and $c'$. Therefore, the term $\gamma H/c'$ cannot be straightforwardly determined. Owing to the implicit nature of the equation, instead of optimizing $\gamma H/c'$, the unit weight $\gamma$, which is function of $H$, $c'$, $\theta_0$, and $\theta_h$ is used as an objective function. For each computation, arbitrary values of $H$ and $c'$ are selected and plugged into Eqs. (12b) and (13b). The unit weight $\gamma$ is minimized with respect to parameters $r_0$ and $r_h$, subject to the following constraints:

(i) $\theta_0 \leq \theta_h$;   (ii) $\theta_0 \geq 0$;   (iii) $\theta_h \geq 0$;   (iv) $L/r_0 \geq 0$;   and (v) $H/r_0 \geq 0$.

Suitable codes are written in *MATLAB 2018a*, and the optimization process is carried out using the FMINCON function on an Intel(R) Core(TM) i5-7200U CPU. Finally, stability numbers are reported as $S_n = \gamma H/c'$ by plugging back the arbitrarily chosen values of $H$ and $c'$. On account of the convexity of the yield surface and linearity of the other constraints, the local minimizer becomes the same as the global minimizer.

## 4. Parametric study

An extensive analysis is carried out to understand how the stability number ($S_n$) gets influenced by the effect of soil types (characterized by soil strength and vG parameters), seismic conditions ($k_h$ and $\lambda = k_v/k_h$), GWT depth ($h_w$), flow conditions ($Q$), and surcharge loadings ($p_s$). The obtained results are elaborately discussed, in turn.

### 4.1. Impact of β and ϕ′ on $S_n$

Figs. 3 and 4 present the variation of $S_n$ with slope angles ($\beta$) and effective friction angle ($\phi'$) corresponding to $k_h$=0.1, and $k_h$=0.3, respectively. The $S_n$ versus $\beta$ and $S_n$ versus $\phi'$ are designated here as $S_n(\beta)$ and $S_n(\phi')$ curves and they are represented with solid and dashed lines, respectively. The analyses in this section are carried out for (a) two different $\lambda$'s ($\lambda$=0 and $\lambda$=1), (b) three different AEVs ($\alpha$=0.1, 0.01 and 0.001 kPa$^{-1}$), and (c) different flow conditions ($Q$ = 0.5, 0.0, and −0.5). The no-suction cases (represented with NS curves), in a sense, represent the stability of the completely saturated soils. The desaturation rate is taken to be the same ($n$=3.0) for all the cases. While plotting $S_n(\beta)$, $\phi'$ is taken as 30°, and for tracking the trend of $S_n(\phi')$ curves, $\beta$ is considered to be 45°. Evidently, $S_n$ decreases with (a) increase in $\beta$, and (b) decrease in $\phi'$. Higher $k_h$ apparently suppresses the stability of the slope. Nevertheless, irrespective of $k_h$, the trend of the stability curves remains almost the same; the slope angle-induced stability-decrement-rate slightly enhances for lower seismic acceleration. However, the variation of $\lambda$ does not seem to be much impactful. This is further explored in the subsequent section. The steady-state flow direction similarly impacts the $S_n(\beta)$ and $S_n(\phi')$ curves, referred here as stability curves. Regardless of the unsaturated properties and the seismic conditions, the evaporation-induced stability curves remain at the top and the stability curves pertaining to the no-suction state becomes the bottommost curves. Below the EV stability curve lies the NF stability curve and that is followed by IF stability curve. It can be interpreted that the evaporative condition yields the highest stability number, and the infiltrative flow results in the least stability number. The stability curves are plotted for various $\alpha$ values. For $\alpha = 0.1$, the $S_n$ values pertain to upward water flow becomes complex number, and hence, the corresponding $S_n$ curves were unattainable. for both the considered $k_h$. Furthermore, the gap between the NF and IF stability curve shrinks for high AEV soils; the reduction of this gap seems to be quite appreciable when $\alpha$ decreases from 0.1 kPa$^{-1}$ to 0.01kPa$^{-1}$, however, this gap is unnoticeable when $\alpha$ decreases from 0.01 kPa$^{-1}$ to 0.001kPa$^{-1}$.

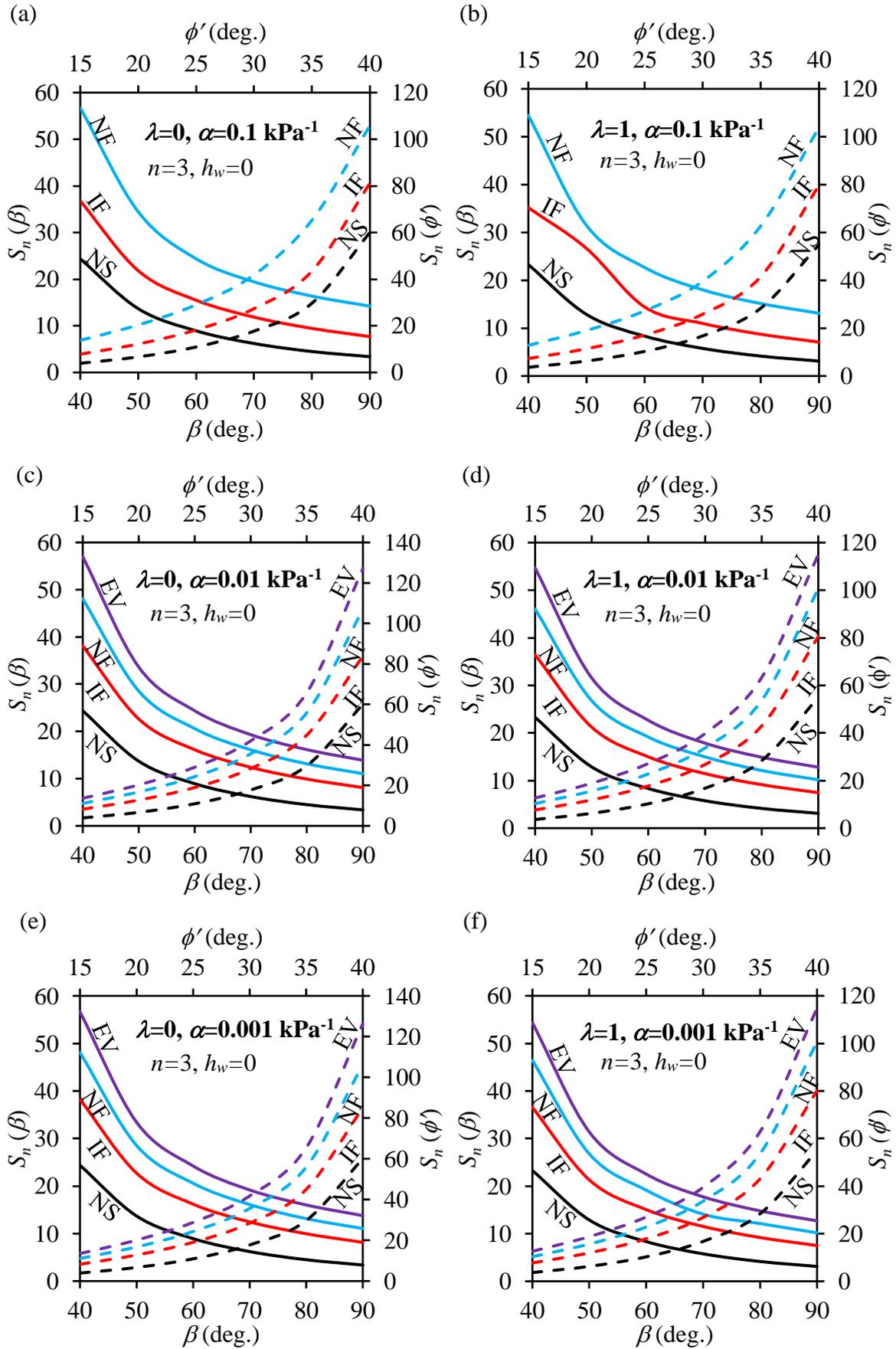

Note: (i) Solid and dashed lines represent $S_n(\beta)$ and $S_n(\phi')$, respectively.
(ii) EV: Evaporation; NF: No-Flow; IF: Infiltration; NS: No Suction

**Fig. 3.** The variation of $S_n$ with $\beta$ and $\phi'$ corresponding to $k_h=0.1$ and $\lambda$ equals to: (a,c,e) 0 and (b,d,f) 1, with three different $\alpha$'s (kPa$^{-1}$): (a,b) $\alpha=0.1$, (c,d) $\alpha=0.01$, and (e,f) $\alpha=0.001$.

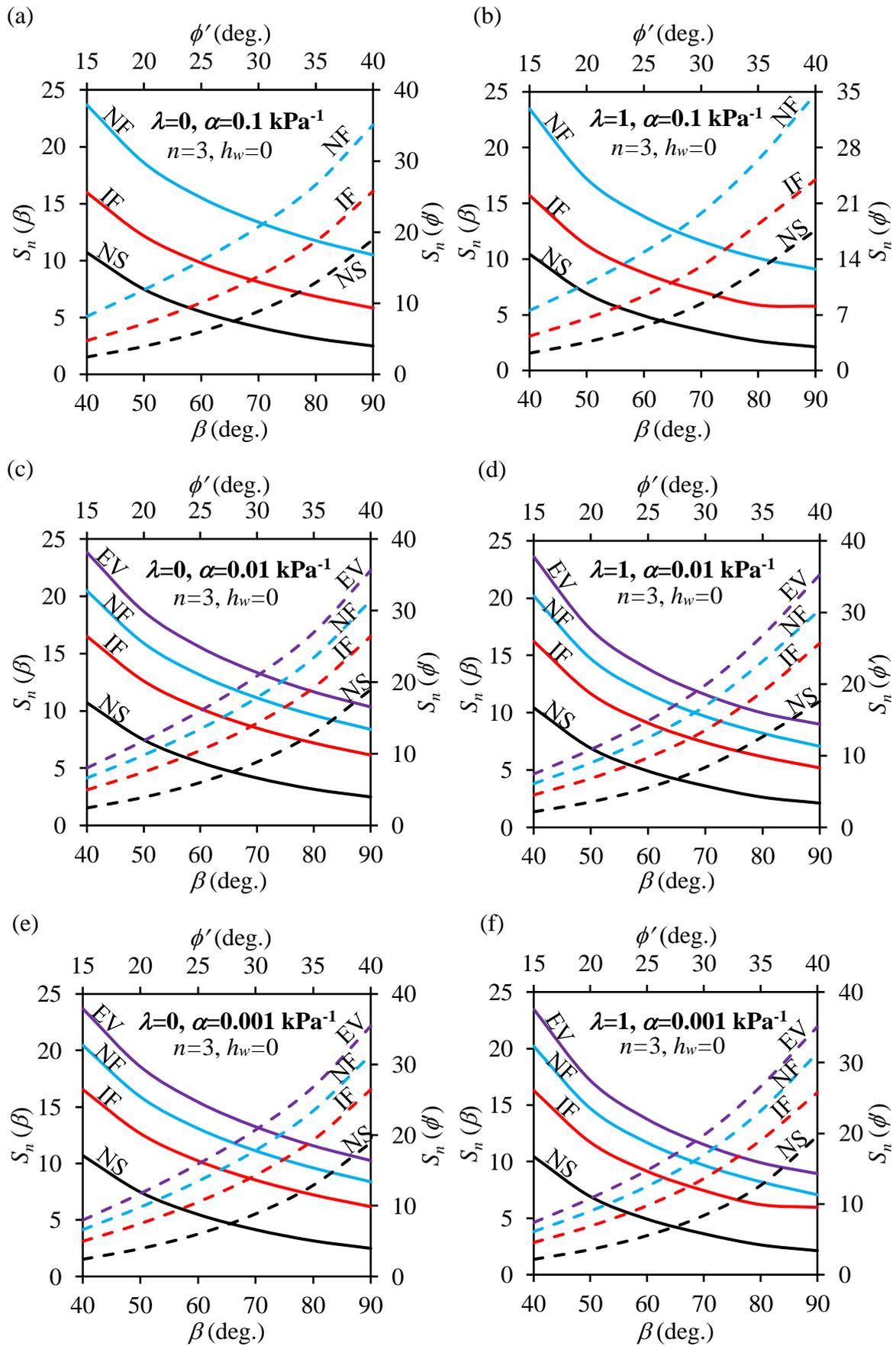

Note: (i) Solid and dashed lines represent $S_n(\beta)$ and $S_n(\phi')$, respectively.
(ii) EV: Evaporation; NF: No-Flow; IF: Infiltration; NS: No Suction

**Fig. 4.** The form of $S_n(\beta)$ and $S_n(\phi')$ curves corresponding to $k_h=0.3$ and $\lambda$ equals to: (a,c,e) 0 and (b,d,f) 1, with three different $\alpha$'s (kPa$^{-1}$): (a,b) $\alpha=0.1$, (c,d) $\alpha=0.01$, and (e,f) $\alpha=0.001$.

## 4.2. Impact of $\lambda$ on $S_n$

Fig. 5 presents the variation of $S_n$ with the parameter $\lambda$ (referred here as $S_n(\lambda)$ curves) by varying the following parameters: (a) flow conditions (i.e. EV, IF, NF, NS), (b) $k_h$ (=0.1 and 0.3) values, (c) $n$ (=2 and 7) parameter, and (d) $\alpha$ (=0.1, 0.01, and 0.001 kPa$^{-1}$) parameter. The analyses are carried out for constant $\phi'$ (=30°) and $\beta$ (=45°). It is observed that the $S_n(\lambda)$ profiles remain almost horizontal indicating that the $\lambda$-value, or in other word the $k_v$ value (for a constant $k_h$) has a negligible impact on the computed stability. Similar to the previous section, $S_n(\lambda)$ corresponding to EV is the topmost curve and it is subsequently followed by the NF, IF, and NS induced $S_n(\lambda)$ curves. The increase in $k_h$ drastically reduces the stability value. The deviation of the $S_n(\lambda)$-profiles corresponding to $k_h$=0.1 and $k_h$ =0.3 remains maximum for the EV flow and minimum for saturation state. There is a visible growth in the deviation between the $S_n(\lambda)$-curves when the soil is poorly-graded ($n$=7) in comparison to its well-graded ($n$=2) counterpart.

## 4.3. Impact of Water Table Depth $h_w$ on $S_n$

Fig. 6 depicts the variation of $S_n$ with $h_w$ (referred here as $S_n(h_w)$ curves) for a wide variety of soil types and water flows, conforming to various: (a) AEVs ($\alpha$=0.1, 0.01, and 0.001 kPa$^{-1}$), (b) $n$ (=2.0, 4.0, and 7.0) values, and (c) flow conditions (EV, NF, IF, and NS). The inclination ($\beta$=45°) and the frictional strength ($\phi$=30°) of the soil are kept to be constant. Apparently, the $S_n(h_w)$ curves for the no-suction (i.e., saturation) condition remain to be the bottommost horizontal curves and unaffected by the $\alpha$ and $n$ parameters. Irrespective of the vG-model parameters, two distinct features are easily perceived: (a) $S_n(h_w)$ curve shows nonlinear increasing trend, and (b) $S_n(h_w)$-NF curve always lies between $S_n(h_w)$-EV and $S_n(h_w)$-IF curves. An exception is being observed for $\alpha$=0.1 kPa$^{-1}$, $n$=2.0 where the $S_n(h_w)$-EV curve displays an inverted 'V'-shape profile; the reason for this abruptness is not well understood. The $S_n(h_w)$-EV curves for high $\alpha$ and $n$ (e.g. $\alpha$=0.1 kPa$^{-1}$, $n$=7.0) are unattainable owing to the imaginary solutions. As the AEV of the soil increases, there is a substantial improvement in the stability number. This improvement seems to be more pronounced when $\alpha$ varies from 0.1 kPa$^{-1}$ to 0.01 kPa$^{-1}$ rather than 0.1 kPa$^{-1}$ to 0.01 kPa$^{-1}$. The influence of $n$ becomes more visible for low AEV soils; the higher the $n$ the higher the stability. Not only the stability magnitude, the vG model parameters also influence the trend of $S_n(h_w)$ curves. For relatively smaller $\alpha$ and $n$ (e.g. $\alpha$=0.1 kPa$^{-1}$, $n$=7.0), the $S_n(h_w)$ curves increase up to a certain GWT position and thereafter it manifests a horizontal plateau. This indicates the existence of a critical water table, especially for the coarse-grained soil, beyond which there is no further improvement of the stability. With the increase in $\alpha$, the $S_n(h_w)$ curves exhibit a continuous increasing trend within the chosen GWT depth. The $n$-value also reduces the curvature of the $S_n(h_w)$ profiles.

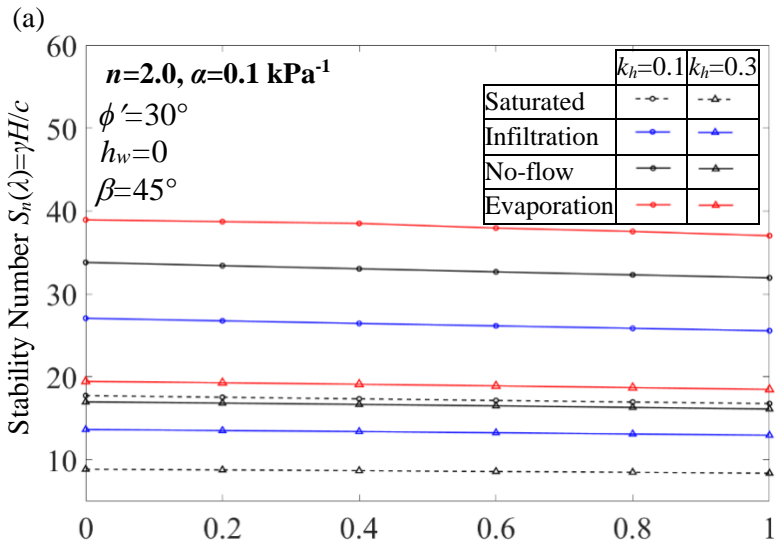
(a)
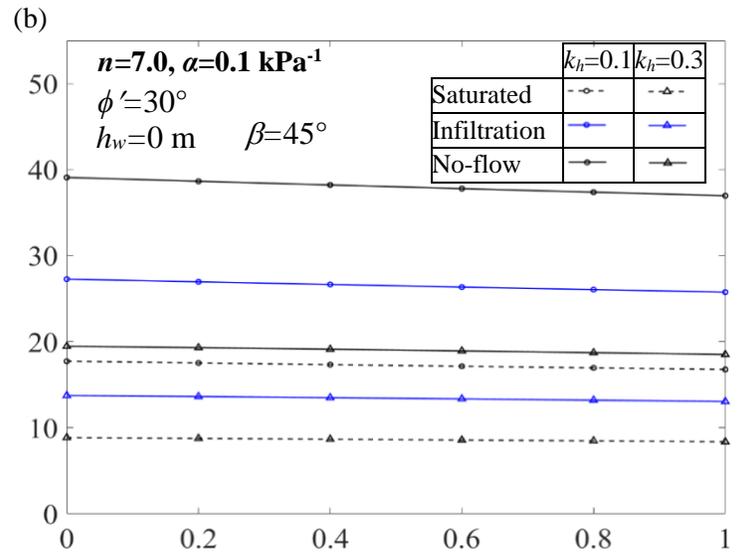
(b)
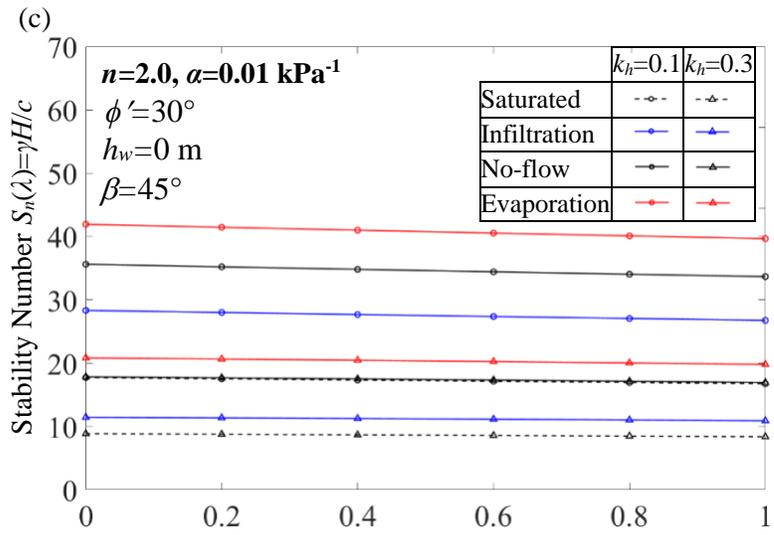
(c)
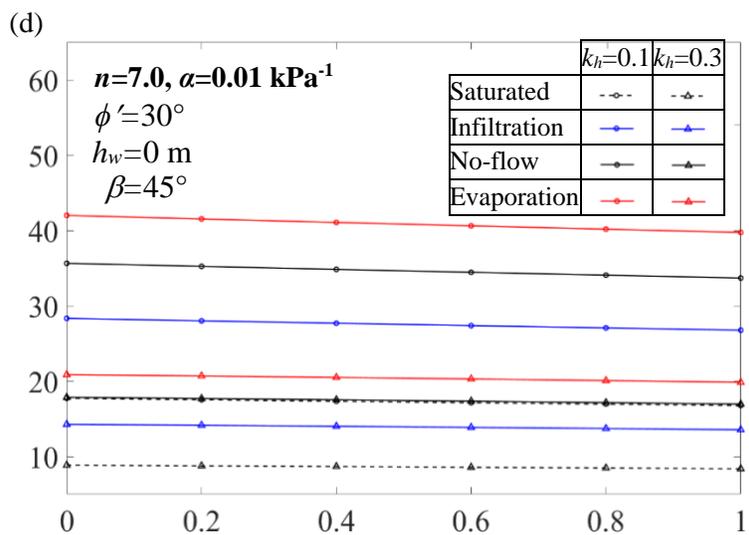
(d)
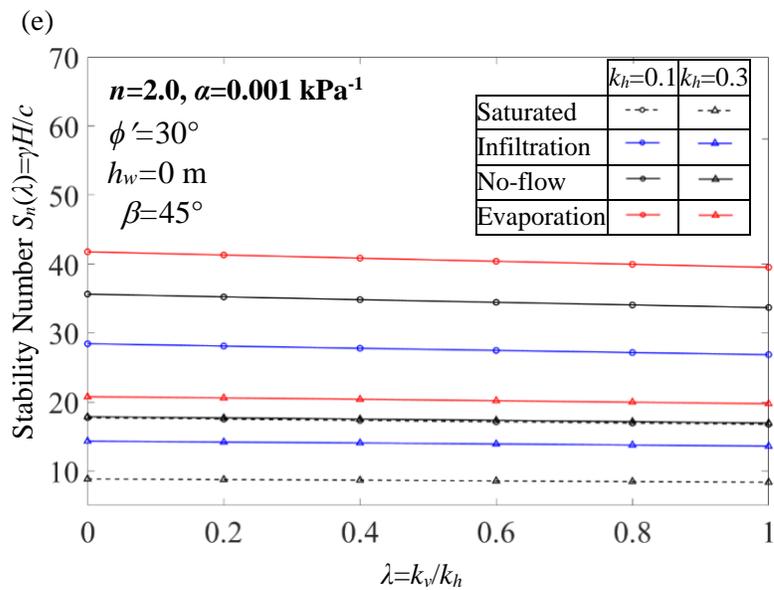
(e)
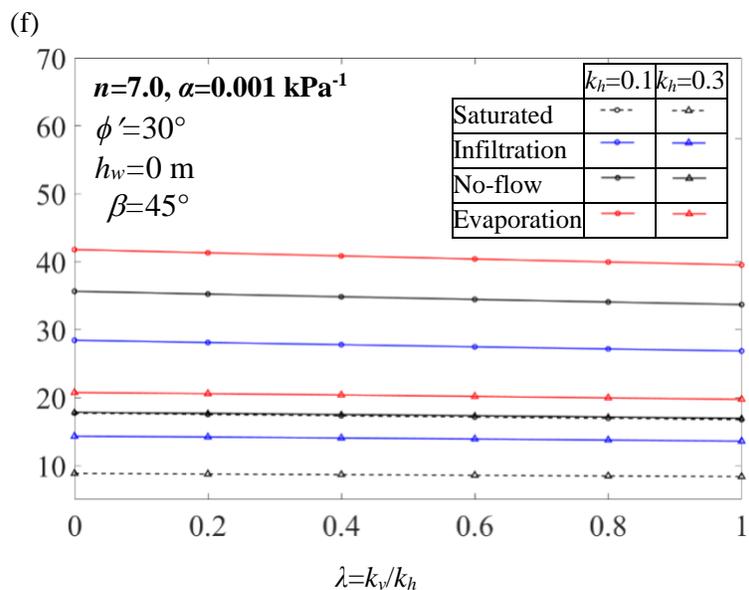
(f)

**Fig. 5.** The form of $S_n(\lambda)$ curves corresponding to various combinations of vG -model parameters: (a) $n=2.0$, $\alpha=0.1$ kPa$^{-1}$, (b) $n=7.0$, $\alpha=0.1$ kPa$^{-1}$, (c) $n=2.0$, $\alpha=0.01$ kPa$^{-1}$, (d) $n=7.0$, $\alpha=0.01$ kPa$^{-1}$, (e) $n=2.0$, $\alpha=0.001$ kPa$^{-1}$, and (f) $n=7.0$, $\alpha=0.001$ kPa$^{-1}$.

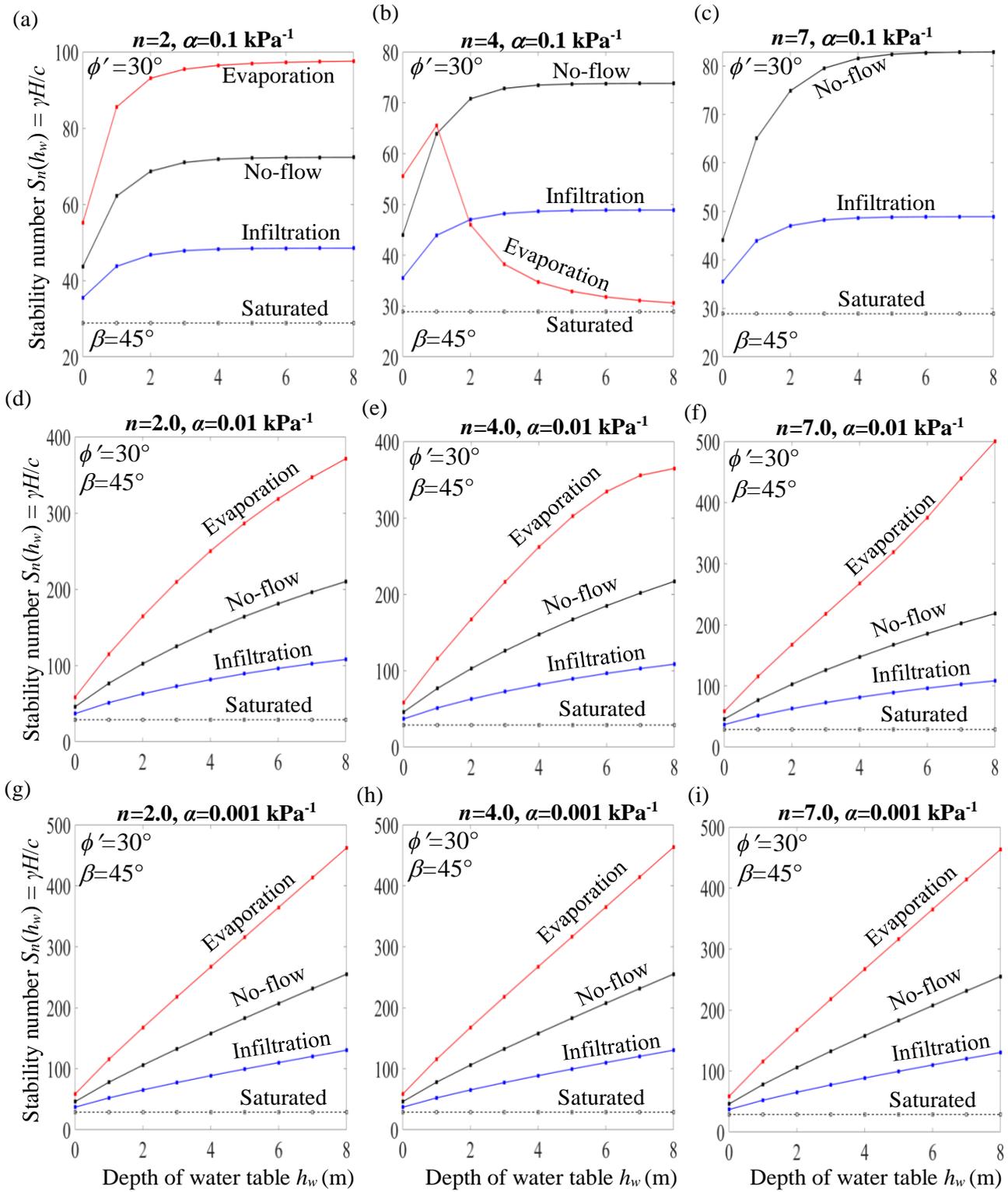

**Fig. 6.** The form of $S_n(h_w)$ curves corresponding to various combinations of vG $n$-parameter: (a, d, g) $n=2.0$, (b,e,h) $n=4.0$, (c,f,i) $n=7.0$ and vG $\alpha$-parameter: (a,b,c) $\alpha = 0.1$ kPa$^{-1}$, (d,e,f) $\alpha = 0.01$ kPa$^{-1}$, and (g,h,i) $\alpha = 0.001$ kPa$^{-1}$.

## 4.4. Combined Impact of $k_h$ and $h_w$ on $S_n$

Fig. 7 depicts the combined effect of horizontal seismic coefficient ($k_h$) and GWT depth ($h_w$) on stability number ($S_n$) for various $\alpha$ and $\lambda$. The three-dimensional curves are plotted for 45° slope without the application of any surcharge pressure. The soil is having a frictional strength of 30° and is subjected to relatively high infiltration ($Q= -0.8$). These three-dimensional surfaces, in a sense, represent the combined effect of the seismic loading and the GWT position that are discussed in the previous sections. The profile seems to be nonlinear wrapped one. Apparently, $S_n$ increases with the lowering of the GWT. This improvement in the stability happens to be more visible for the static loading rather than the pseudo-static one. The combined influence of soil's air-entry value, GWT position, and the horizontal and vertical seismic forces are clearly discernible from the obtained three-dimensional surfaces.

## 4.5. Impact of Surcharge Load ($p_s$) on $S_n$

Fig. 8 depicts the variation of $S_n$ with $p_s$ (referred here as $S_n(p_s)$ curves) for a wide variety of vG parameters conforming to two different friction angles, namely $\phi'=20°$ and $\phi'=35°$. The simulations were carried out by considering $\beta = 50°$, $h_w=2$m, $Q=-0.3$. For the verification of $n$ parameter on $S_n(p_s)$ curves, $\alpha$ is taken to be 0.05 kPa$^{-1}$, whereas for investigating the impact of $\alpha$ parameter on $S_n(p_s)$ curves, the magnitude of $n$ is adopted as 6. No matter whatsoever be the unsaturated properties, $S_n(p_s)$ curves always manifests a decreasing trend. This can be attributed to the fact that the surge in surcharge pressure further intensifies the destabilising force. For $\phi'=20°$, the n parameter hardly impacts the $S_n(p_s)$ curves. Nevertheless, the soils with relatively higher strength ($\phi'=35°$) exhibit noticeable influence on the $S_n(p_s)$ curves; the $S_n(p_s)$ curve pertains to lower $n$, remains in the bottom side. Conversely, the vG-$\alpha$ parameter plays a dominating role in dictating the form of the $S_n(p_s)$ curves. The changes in $S_n(p_s)$ profiles are highly observable when $\alpha$ changes from 0.01 kPa$^{-1}$ to 0.1 kPa$^{-1}$; the higher the $\alpha$, the lower the stability values. The decreasing trends of the $S_n(p_s)$ curves clearly reveal the presence of certain $p_s$ at which $S_n$ will attain zero magnitude; this certain $p_s$ is designated here with the symbol $p_{sc}$. The magnitude of $p_{sc}$ becomes very less for a weaker soil, especially having lower AEV. For a stronger soil (e.g. $\phi'=35°$) the attainment of $p_{sc}$ happens at a very large surcharge pressure, because the decrement rate of $S_n(p_s)$ curves occurs at a relatively slower rate. Fig. 9 depicts the three-dimensional stability chart ($S_n$-$p_s$-$\alpha$) for a 45° (=$\beta$) soil slope having $\phi'=25°$ and subjected to evaporation ($Q=-0.4$); Figs. 9a and b correspond to $h_w=0$ m and $h_w=4$m, respectively. The escalated surcharge pressure minimizes the nonlinear relation between $S_n$ and $\alpha$. Furthermore, with the increase in $h_w$, the occurrence of $p_{sc}$ happens much earlier.

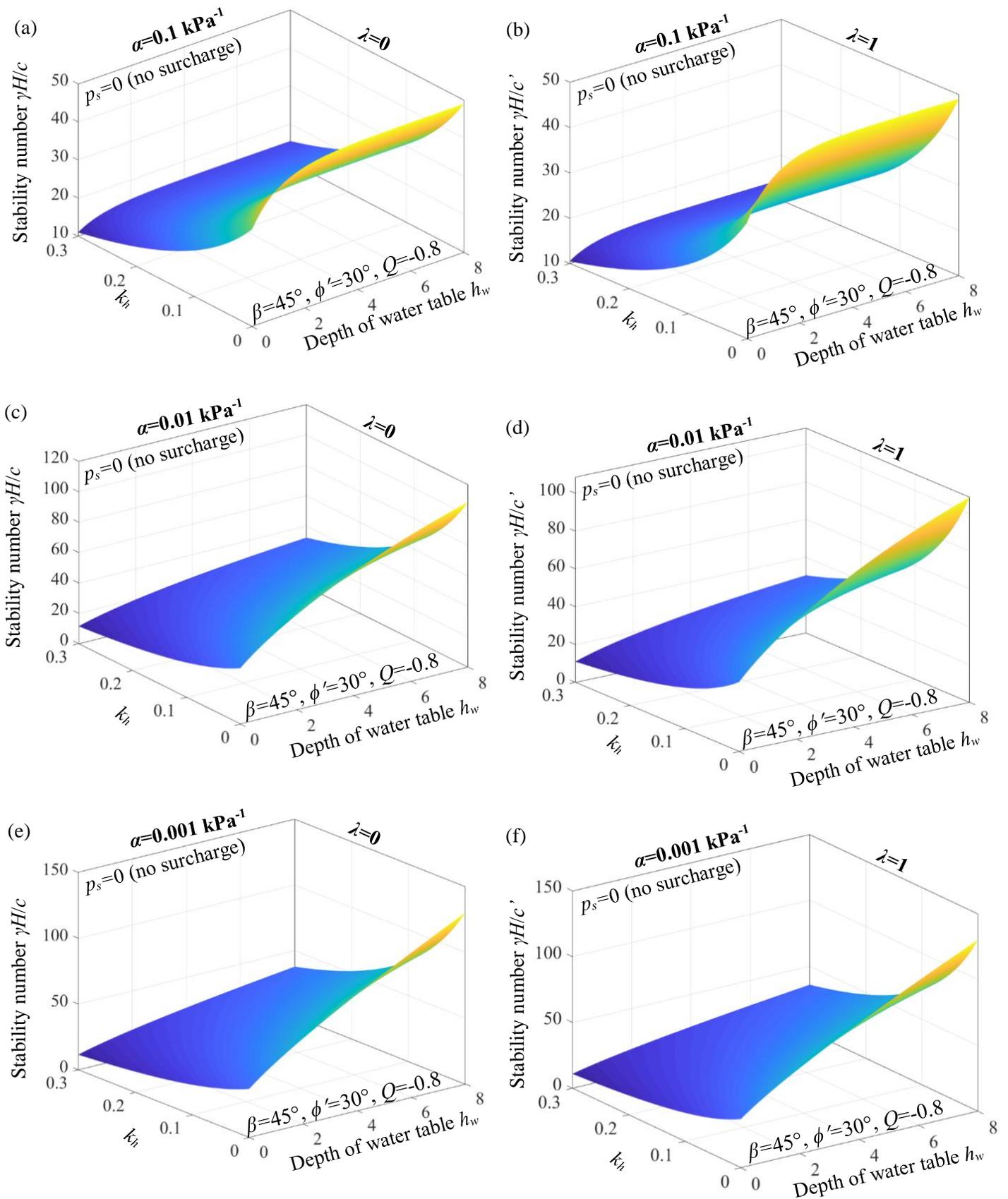

**Fig. 7.** The three-dimensional stability profiles in $S_n - k_h - h_w$ space for three different $\alpha$'s: (a,b) $\alpha = 0.1$ kPa$^{-1}$, (c,d) $\alpha = 0.01$ kPa$^{-1}$, and (e,f) $\alpha = 0.001$ kPa$^{-1}$ and two $\lambda$'s (a,c,e) $\lambda = 0$ and (b,d,f) $\lambda = 1$.

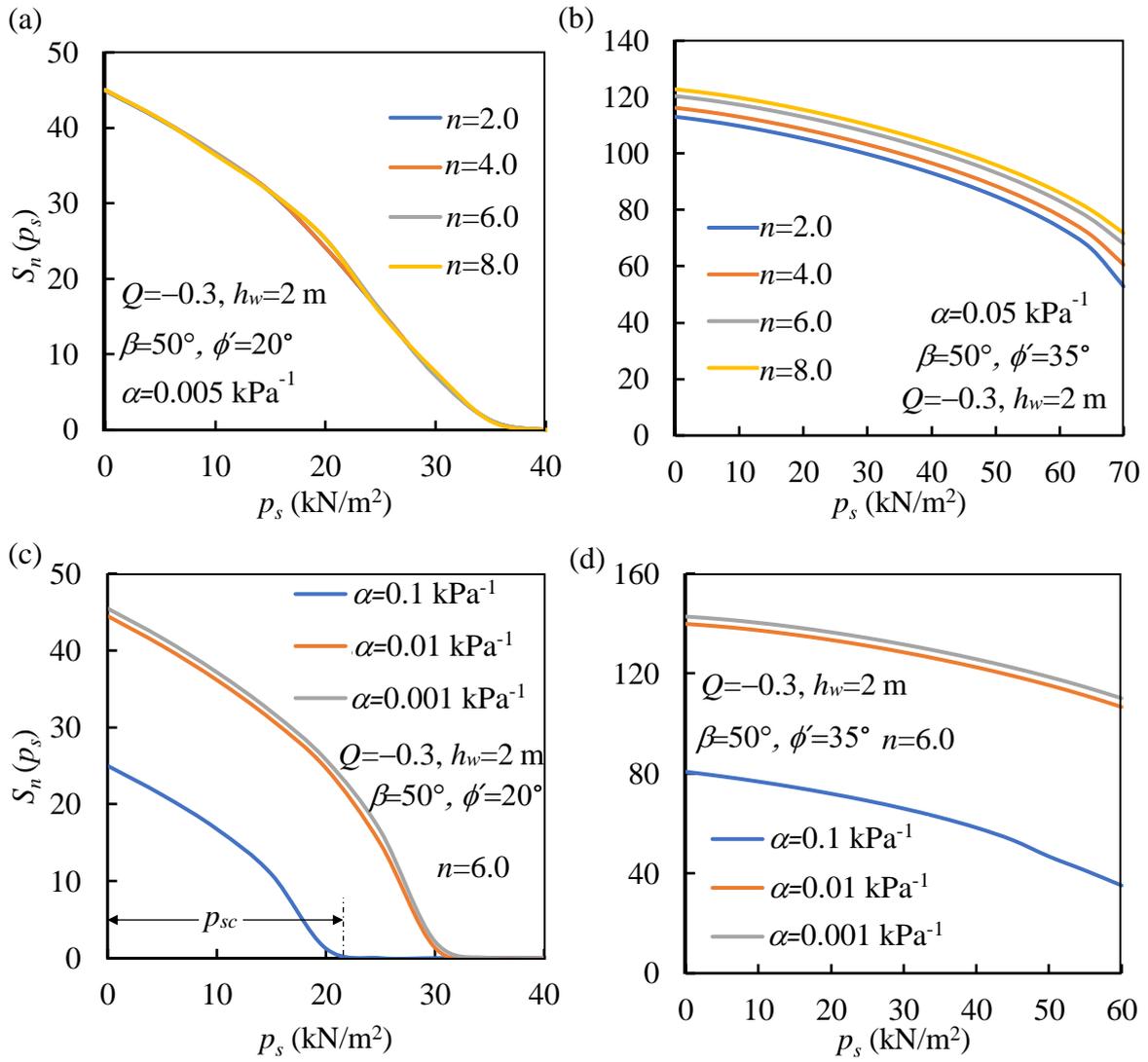

**Fig. 8.** Variation of $S_n(p_s)$ with surcharge load $p_s$ with respect to (a) and (b) vG parameter $n$; (c) and (d) vG parameter $\alpha$.

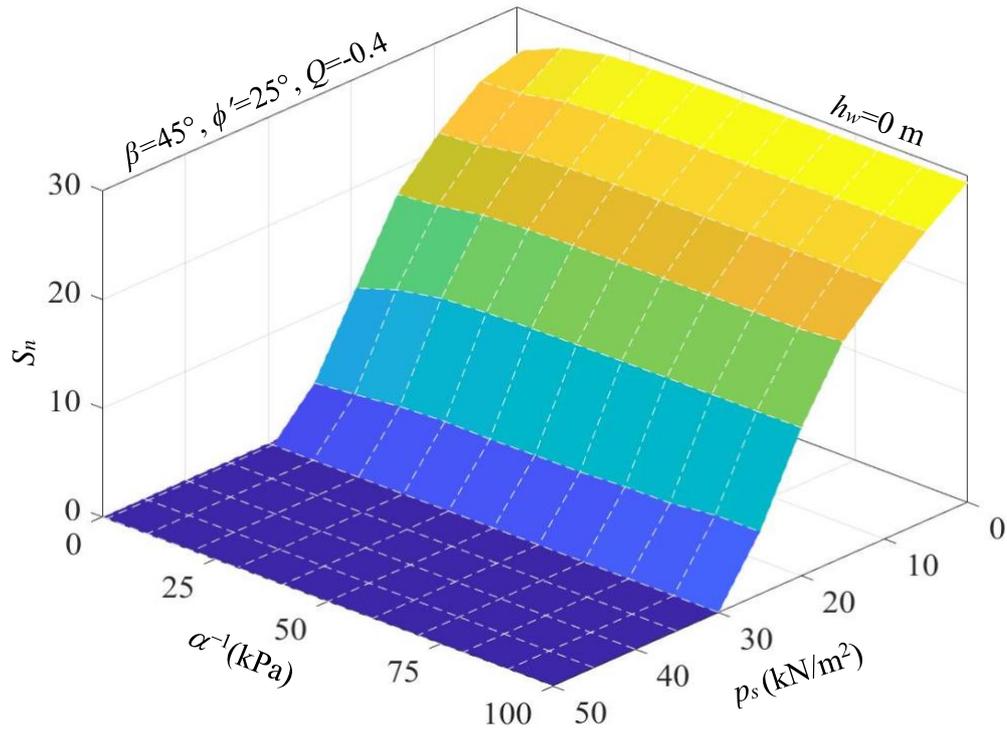

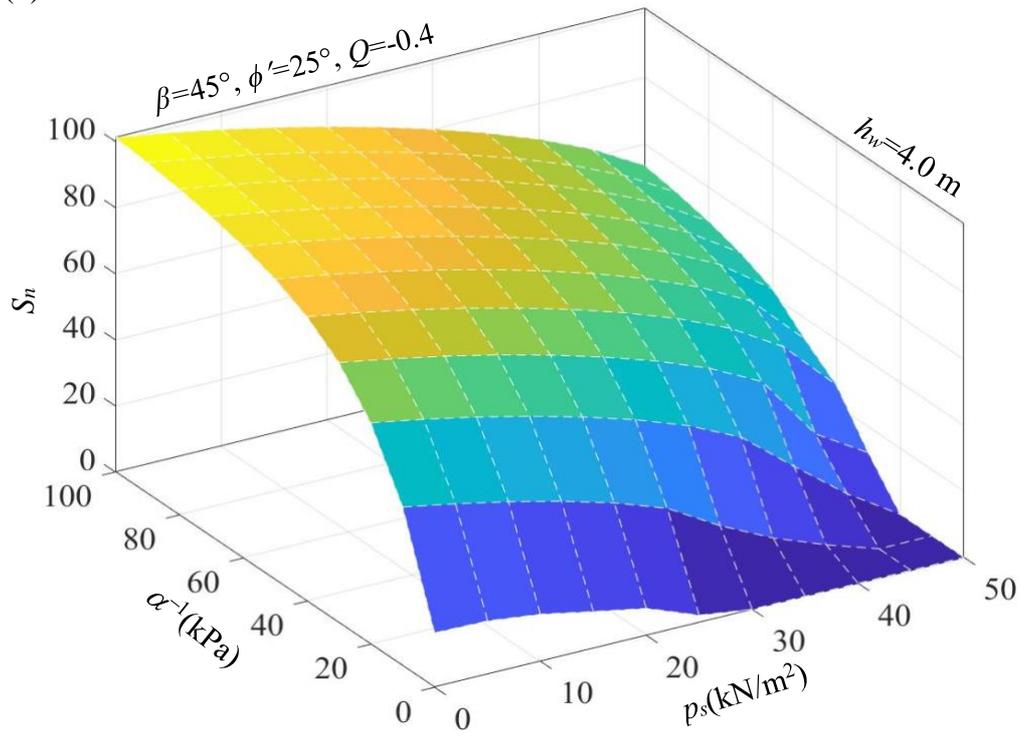

**Fig. 9.** Three dimentional represention of the variation of $S_n$ with vG parameter $\alpha$ and $p_s$ conforming to the parameters: $\phi'=25°$, $Q=-0.4$, $\beta=45°$ for (a) $h_w = 0$ m and (b) $h_w = 4.0$ m.

## 5. Failure surfaces

Fig. 10 presents a comparative study of the failure surfaces generated for a representative slope ($\beta=45°$, $\phi'=20°$, and $h_w=0$) under hydrostatic condition. The failure surfaces are plotted for different seismic condition ($k_h$=0.0, 0.1, 0.2, and 0.3; $\lambda$=0, 1) and different air-entry values ($\alpha$= 0.1, 0.01, and 0.001 kPa$^{-1}$). It is observed that the log spiral failure surface shifts to the right as the value of $k_h$ increases from $k_h$=0.0 to $k_h$=0.3. This can be interpreted as the fact that the volume of soil encompassed within the failure zone expands with the seismic effects. The gap between the failure surfaces corresponding to various $k_h$ becomes more visible for $\lambda$=0 than $\lambda$=1.

Fig. 11 depicts the impact of the surcharge pressure and the GWT depth on the trend and extent of the failure surfaces for specific soil slopes ($\beta=45°$, $\phi'=35°$, $n=4$, and $\alpha$= 0.1/ 0.005 kPa$^{-1}$); Figs. 11a and b displays the failure surfaces by varying $p_s$ whereas, Figs. 11c and d displays the failure surfaces by varying the GWT depth. The surcharge pressures contribute to the expansion of soil volumes enclosed by failure surfaces. The curvature of the failure surfaces reduces with the increase in surcharge pressure. For high AEV soils, the failure surfaces coincide adjacent to the toe zones; these surfaces become discernible well above the slope toe, and the most significant deviation observed at the ground surface where $p_s$ is acting. Nevertheless, soil featuring lower AEV, tends to exhibit distinctive failure surfaces even near the toe zone of the slopes. The fluctuation of GWT below the slope toe also plays a crucial role in shaping the development of the failure surface. The deeper the GWT, the less extensive the failure zone becomes. Beyond a certain GWT depth, the failure surface does not undergo any further alterations.

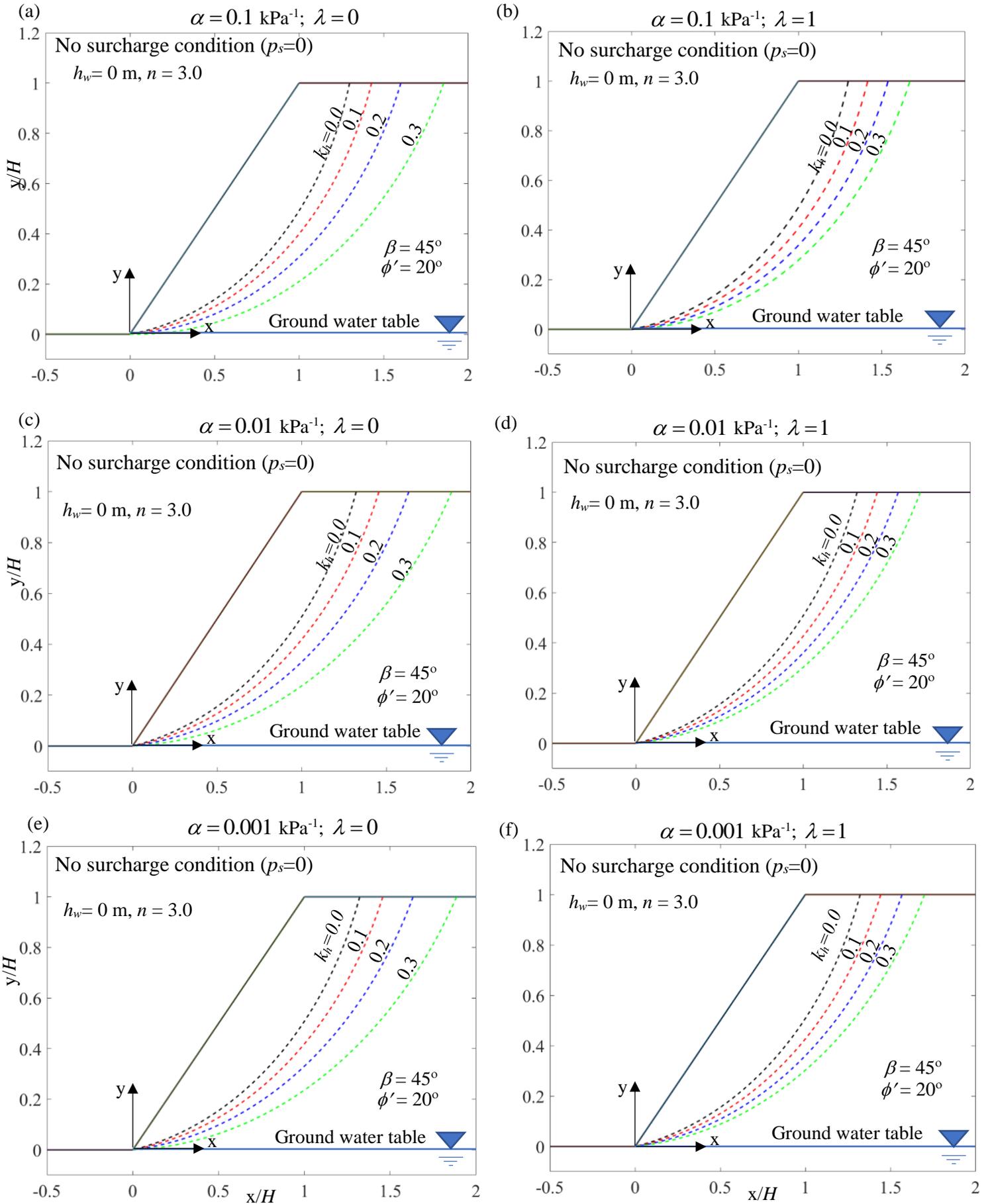

**Fig. 10.** The form of the developed failure surfaces corresponding to $\beta = 45°$, $\phi' = 20°$, $h_w = 0$ m, $n = 3.0$ and conforming to (a) $\alpha$=0.1 kPa$^{-1}$; $\lambda$=0, (b) $\alpha$=0.1 kPa$^{-1}$; $\lambda$=1, (c) $\alpha$=0.01 kPa$^{-1}$; $\lambda$=0, (d) $\alpha$=0.01 kPa$^{-1}$; $\lambda$=1, (e) $\alpha$=0.001 kPa$^{-1}$; $\lambda$=0, (f) $\alpha$=0.001 kPa$^{-1}$; $\lambda$=1.

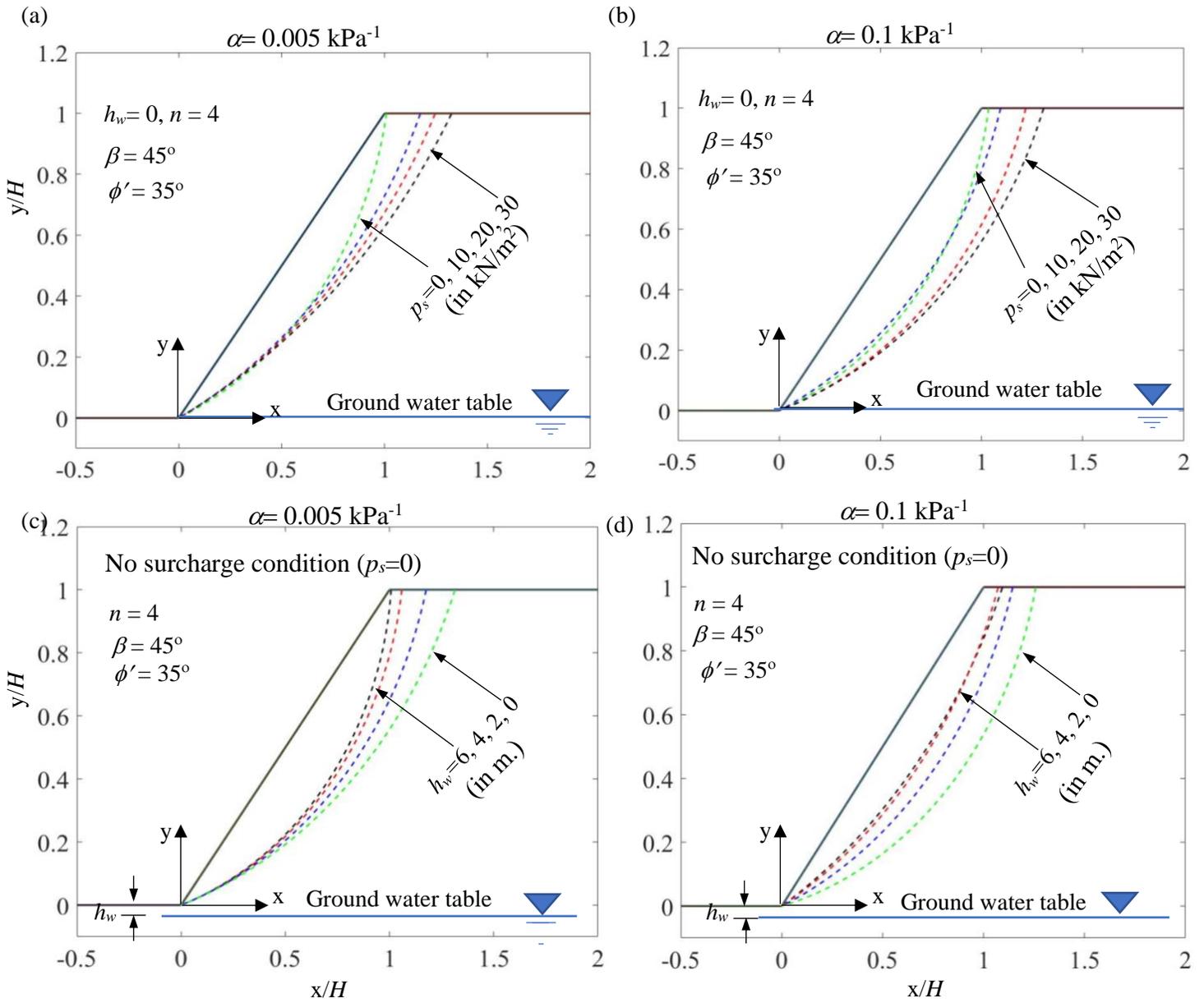

**Fig. 11.** The failure surfaces by varying: (a, b) $p_s$ and (c, d) GWT depth for two different $\alpha$'s: (a, c) 0.005 kPa$^{-1}$ and (b, d) 0.1 kPa$^{-1}$.

# 6. Verification with literature

To verify the efficacy and the accuracy of the proposed methodology, the failure surfaces obtained from the present analysis are compared with that of Vahedifard et al. (2016) for a representative soil slope pertaining to non-seismic, surcharge-less conditions and having $\beta=70°$, $\phi=25°$. Fig. 12 depicts such comparison for two characteristic cases, namely, no suction (NS), and unsaturated soil with no flow ($\alpha=0.005$ kPa$^{-1}$, $n=2.0$, $Q=0$). The obtained failure surfaces are in good agreement with the solutions of Vahedifard et al. (2016), especially for NS case.

Further, Fig. 13 shows the comparison of the $S_n(\phi')$ profiles obtained from the present study with the results of Vahedifard et al. (2016) and Sun et al. (2019) obtained through limit equilibrium framework. The analyses are performed on two different vertical unsaturated soil ($\alpha=0.005$ kPa$^{-1}$, $n=2.0$, and $\alpha=0.1$ kPa$^{-1}$, $n=5.0$) slopes that are subjected to no-flow, non-seismic, surcharge-less conditions. It is noteworthy that the Taylor's stability number reported by Vahedifard et al. (2016) and Sun et al. (2019) are inverted for comparing the present solution documented in the form of Chen's stability number ($S_n$). The NS envelopes evolved from the present analysis almost coincide with its counterpart solutions of Vahedifard et al. (2016) and Sun et al. (2019). Evidently, in presence of the suction stress, the stability of the soil improves. As the slope height increases, the impact of suction stress diminishes and the $S_n(\phi')$ envelopes move closer to the NS curve. This observation is in good accordance with the available literatures. However, the present solutions seem to be quite conservative, especially for higher friction values. Unlike the sudden jerk of the $S_n(\phi')$ profiles, as reported in the previous literatures, the present analysis leads to smooth and monotonic increasing trends of the $S_n(\phi')$ envelopes, providing a more realistic representation of slope stability.

Fig. 14 presents the comparative variation of $S_n(\lambda)$ profiles with the upper bound solutions of Zhao et al. (2016) for a surcharge less saturated soil slope ($\beta=45°$; $\phi'=20°$) subjected to various level of seismicity, $k_h=0.1$, $k_h=0.2$, and $k_h=0.3$. The trend observed from the present analysis matches quite well with Zhao et al. (2016). Notably, the present analysis yields lower $S_n(\lambda)$ values (indicating better upper-bound solutions) compared to Zhao et al. (2016). This numerical disparity may also be partly attributed to Zhao et al.'s (2016) consideration of tensile cracks, a factor not accounted for in this study.

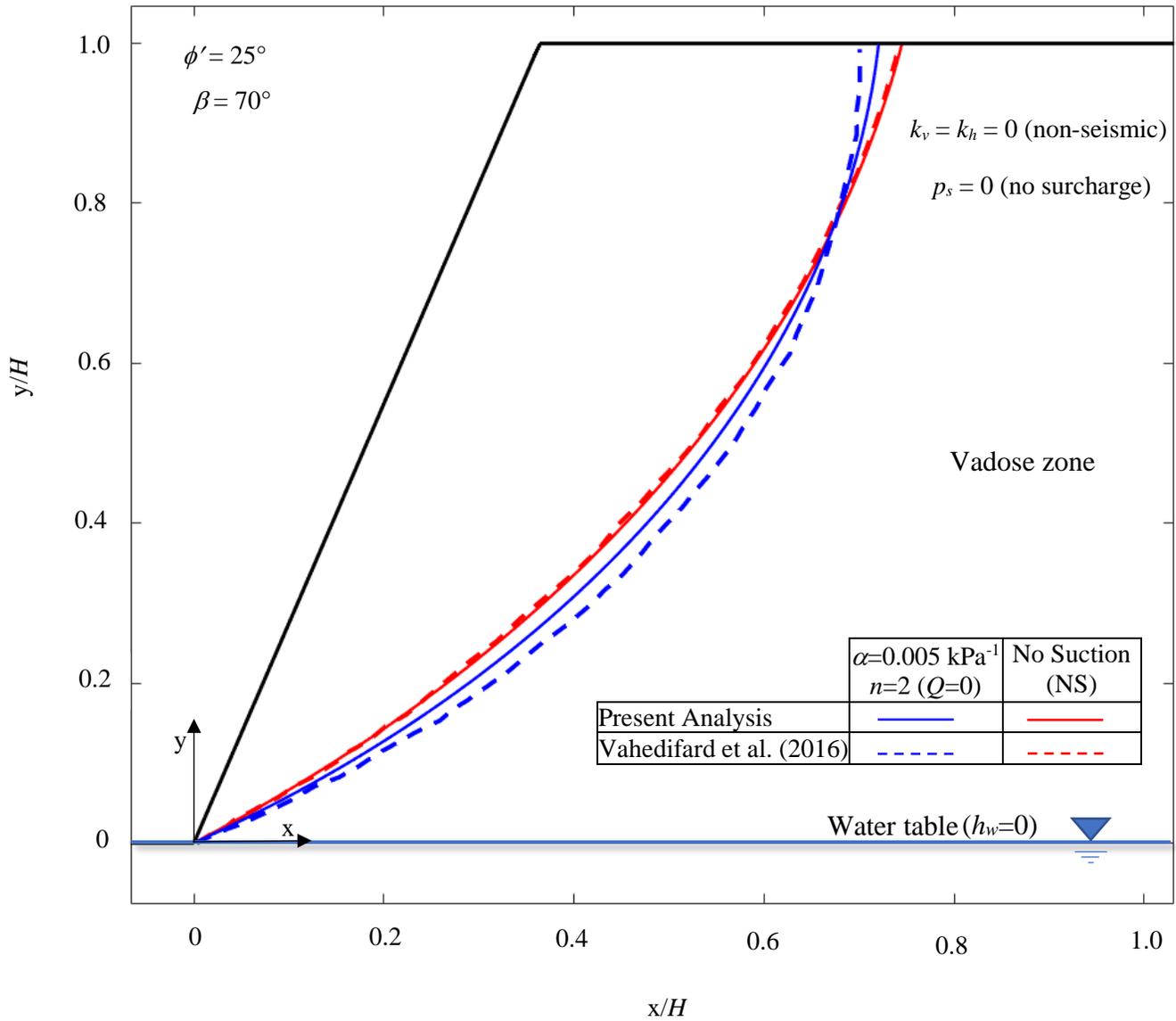

**Fig. 12.** Log-spiral failure surfaces generated for two soil types, namely, an unsaturated soil ($\alpha$=0.005 kPa$^{-1}$, $n$=2) and a saturated soil (no suction) for the representative slope with geometric parameters: $\beta$=70°, $\phi'$=25°. The surfaces generated with the present analysis are compared with that of Vahedifard et al. (2016).

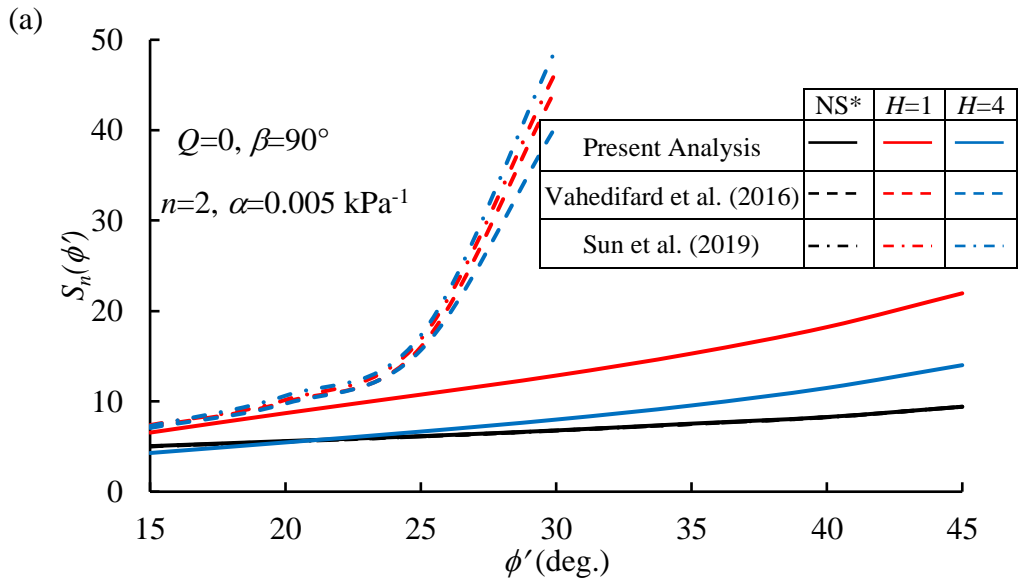

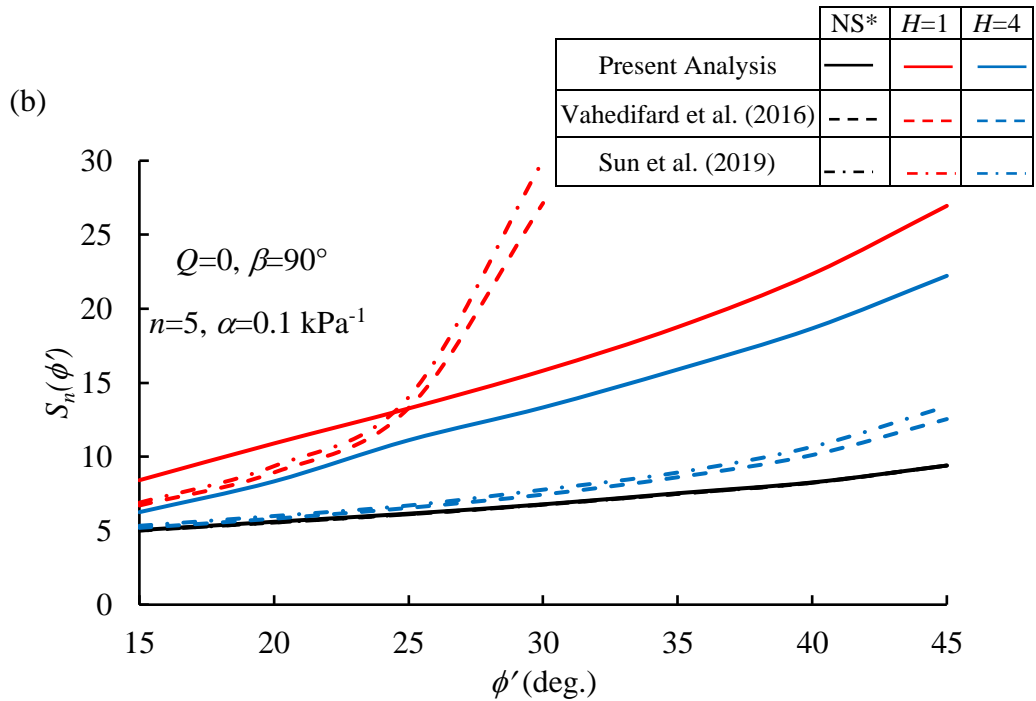

**Note:** *NS= No Suction

**Fig. 13.** A comparison of stability number $S_n(\phi')$ obtained from present study with the solution of Sun et al. (2019) and Vehedifard et al. (2016) considering (a) $\alpha=0.005$ kPa$^{-1}$ and (b) $\alpha=0.1$ kPa$^{-1}$.

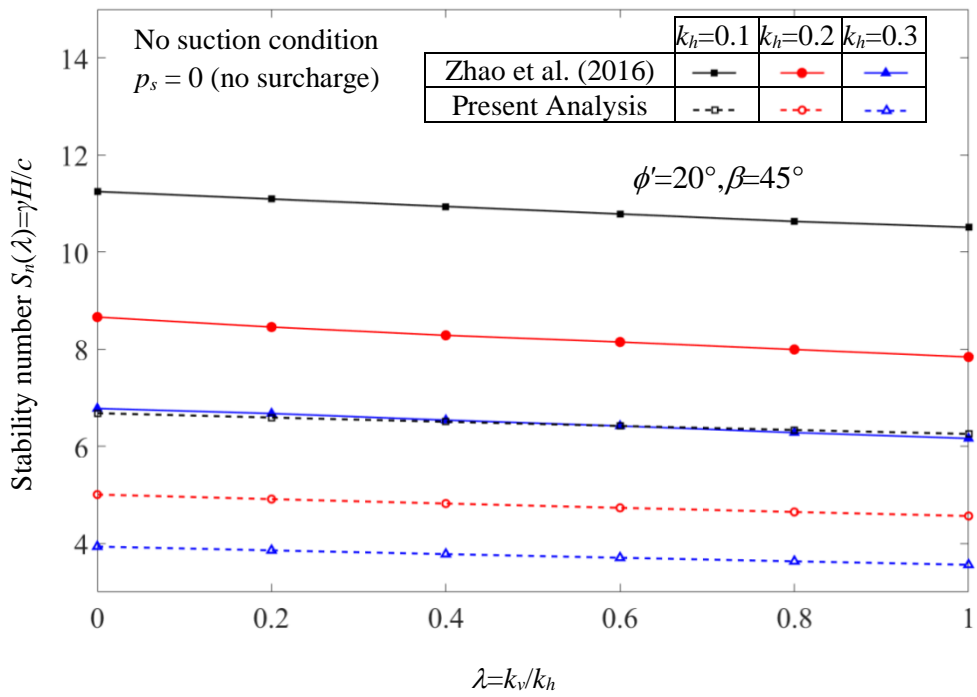

**Fig. 14.** A comparison of stability number obtained from present study and the solution of Zhao et al. (2016) for different $k_h$.

# 7. Conclusions

This article focuses on the computation of the stability number ($S_n$) for homogeneous unsaturated slopes under seismic and surcharge loads. The unified suction-stress-based effective stress formulation is employed to simulate the stability problem by considering van-Genuchten's SWCC model, Gardner's one-parameter HCF, Darcy's linear steady-state flow, rigid plasticity, coaxiality, and associativity. A log-spiral failure mechanism is chosen and an energy balance equation is employed for obtaining the upper bound stability number. To simplify the complexities involved in the dynamic analysis, of earthquake forces, the pseudo-static analysis is engrained into the formulations by factoring in the seismic acceleration coefficients. A thorough parametric analysis is conducted by varying the slope geometry, surcharge pressure, soil's hydromechanical properties, groundwater depth, seismic loadings. The use of an implicit optimization technique improves the computation accuracy and efficiency, representing a significant advancement. The obtained solutions are presented in two-dimensional stability charts $S_n(\beta)$, $S_n(\phi')$, $S_n(\lambda)$, $S_n(h_w)$, and $S_n(p_s)$ for various air entry values and desaturation rates. The combined effects of the chosen parameters are also displayed in three-dimensional surfaces ($S_n$–$h_w$–$k_h$, and $S_n$–$\alpha$–$p_s$). The developed failure contours for different combinations provide valuable perspectives insights into slope behaviour. The computed solutions are validated through a comparison with available literature, showing good agreement and reinforcing the reliability of the proposed methodology. The proposed stability charts provide guidance to engineers, enabling the assessment of slope stability in homogeneous unsaturated soil under seismic and surcharge loading conditions.

# Acknowledgments

The corresponding author acknowledges the support of "Science and Engineering Research Board (SERB), Government of India" under grant number SRG/2019/000149.

# References

[1] Cai, F., Ugaim K., 2004. Numerical analysis of rainfall effects on slope stability. Int. J. Geomech. ASCE 4(2): 69–78.

[2] Chakraborty, M., Kumar J., 2015. Bearing capacity factors for a conical footing using lower-and upper-bound finite elements limit analysis. Canadian Geotechnical Journal, 52(12): 2134-2140.

[3] Chen, T., Xiao, S., 2020. An upper bound solution to undrained bearing capacity of rigid strip footings near slopes. International Journal of Civil Engineering. 18: 475-485.

[4] Chen, W. F., 1975. Limit analysis and soil plasticity. New York, NY, USA: Elsevier.

[5] Cho, S. E., Lee, S. R., 2002. Evaluation of surficial stability for homogeneous slopes considering rainfall characteristics. J. Geotech. Geoenviron. Eng. ASCE 128(9): 756–763.

[6] Cho, S.E., Lee, S.R., 2001. Instability of unsaturated soil slopes due to infiltration. Computers and Geotechnics. 28(3):185-208.

[7] Choudhury, D., Ahmad S.M., 2007. Stability of waterfront retaining wall subjected to pseudo-static earthquake forces. Ocean Eng. 34(s14–15):1947–54.

[8] Costa, Y.D., Cintra, J.C., Zornberg, J.C., 2003. Influence of matric suction on the results of plate load tests performed on a lateritic soil deposit. Geotech. Test. J. 2 (2): 219–226.

[9] Darcy, H., 1856. Les fontaines publiques de Dijon.

[10] Fredlund, D.G., Rahardjo, H., 1993. Soil mechanics for unsaturated soils. John Wiley & Sons.

[11] Gardner, W.R., 1958. Steady state solutions of the unsaturated moisture flow equation with application to evaporation from a water table. Soil Sci. 85: 228–232.

[12] Gavin, K., Xue, J., 2010. Design charts for the stability analysis of unsaturated soil slopes. Geotech. Geol. Eng. 28(1): 79–90.

[13] Griffiths, D.V., Lu, N., 2005. Unsaturated slope stability analysis with steady infiltration or evaporation using elastoplastic finite elements. Int. J. Numer. Anal. Meth. Geomech. 29(3): 249–267.

[14] Hamdhan, I.N., Schweiger, H.F., 2013. Finite element method–based analysis of an unsaturated soil slope subjected to rainfall infiltration. Int. J. Geomech. ASCE 13(5): 653–658.

[15] Kang, S., Lee, S.R., Cho, S.E., 2020. Slope stability analysis of unsaturated soil slopes based on the site-specific characteristics: a case study of hwangryeong mountain, Busan, Korea. Sustainability. 12(7): 2839.

[16] Khalili, N., Khabbaz, M.H., 1998. A unique relationship for $\chi$ for the determination of the shear strength of unsaturated soils. Geotechnique. 48(5): 681-687.

[17] Kumar, J., Chakraborty, M., 2014. Upper-bound axisymmetric limit analysis using the Mohr-Coulomb yield criterion, finite elements, and linear optimization. Journal of Engineering Mechanics. 140(12): 06014012.

[18] Kumar, J., Kouzer, K.M., 2008. Vertical uplift capacity of horizontal anchors using upper bound limit analysis and finite elements. Canadian Geotechnical Journal. 45(5): 698-704.

[19] Kumar, J., Samui, P., 2006. Stability determination for layered soil slopes using the upper bound limit analysis. Geotechnical & Geological Engineering. 24: 1803-1819.

[20] Lu, N., Godt, J., 2008. Infinite slope stability under steady unsaturated seepage conditions. Water Resour. Res. 44(11): W11404.

[21] Lu, N., Likos, W.J., 2004. Unsaturated soil mechanics. Wiley, Hoboken.

[22] Lu, N., Godt, J.W., Wu, D.T., 2010. A closed-form equation for effective stress in unsaturated soil. Water Resour. Res. 46 (5).

[23] Lu, N., Griffiths, D., 2004. Profiles of steady-state suction stress in unsaturated soils. J. Geotech. Geoenviron. Eng. 130 (10). 1063–1076.

[24] Mualem, Y., 1976. A new model for predicting the hydraulic conductivity of unsaturated porous media. Water resources research. 12(3): 513-522.

[25] Oh, W.T., Vanapalli, S.K., 2011. Modelling the applied vertical stress and settlement relationship of shallow foundations in saturated and unsaturated sands. Can. Geotech. J. 48 (3): 425–438.

[26] Oh, W.T., Vanapalli, S.K., 2013. Interpretation of the bearing capacity of unsaturated fine-grained soil using the modified effective and the modified total stress approaches. Int. J. Geomech. 13 (6): 769–778.

[27] Oloo, S.Y., Fredlund, D.G., Gan, J.-K.-M., 1997. Bearing capacity of unpaved roads. Can. Geotech. J. 34 (3): 398–407.

[28] Prasad, S. D., Chakraborty, M., 2023. Upper Bound Collapse Load of Strip Footing that Rests on Unsaturated Sands. International Journal of Geomechanics. 23(4): 04023016.

[29] Roy, S., Chakraborty, M., 2023. Unsaturated bearing capacity of strip foundations by using the upper bound rigid block method. Computers and Geotechnics. 156: 105260.

[30] Singh, V.P., 1997. Kinematic wave modeling in water resources: Environmental hydrology. John Wiley & Sons.

[31] Sloan, S.W. Kleeman, P.W., 1995. Upper bound limit analysis using discontinuous velocity fields. Computer methods in applied mechanics and engineering. 127(1-4): 293-314.

[32] Soubra, A., 1999. Upper-bound solutions for bearing capacity of foundations. J. Geotech. Geoenviron. Eng. 125 (1): 59–68.

[33] Sun D. A., Wang, L., Li, L., 2019. Stability of unsaturated soil slopes with cracks under steady-infiltration conditions. Int. J. Geomech. ASCE 19(6): 04019044.

[34] Sun D. M., Li, X.M., Feng, P., Zang, Y., 2016. Stability analysis of unsaturated soil slope during rainfall infiltration using coupled liquid-gas-solid three-phase model. Water Sci. Eng. 9(3): 183–194.

[35] Thota, S.K. Vahedifard, F., 2021. Stability analysis of unsaturated slopes under elevated temperatures. Engineering Geology. 293:106317.

[36] Travis, Q.B., Houston, S.L., Marinho, F.A., Schmeeckle, M., Unsaturated infinite slope stability considering surface flux conditions. J. Geotech. Geoenviron. Eng. ASCE 136(7): 963–974.

[37] Utili, S., 2013. Investigation by limit analysis on the stability of slopes with cracks. Geotechnique. 63(2): 140-154.

[38] Vahedifard, F., Leshchinsky, D., Mortezaei, K., Lu, N., 2016. Effective stress-based limit-equilibrium analysis for homogeneous unsaturated slopes. Int. J. Geomech., ASCE 16(6): D4016003.

[39] Vahedifard, F., AghaKouchak, A., Robinson, J.D., 2015a. Drought threatens California's levees. Science 349 (6250): 799.

[40] Vahedifard, F., Leshchinsky, B., Mortezaei, K., Lu, N., 2015b. Active Earth Pressures for Unsaturated Retaining Structures. J. Geotech. Geoenviron. Eng. 141 (11): 04015048.

[41] van Genuchten, M.T., 1980. A closed-form equation for predicting the hydraulic conductivity of unsaturated soils. Soil Sci. Soc. Am. J. 44: 892–898.

[42] Vanapalli, S.K., Mohamed, F.M.O., 2007. Bearing capacity of model footings in unsaturated soils. Springer Proceed. Phys. 112: 483–493.

[43] Zhao, L.H., Cheng, X., Zhang, Y., Li, L., Li, D.J., 2016. Stability analysis of seismic slopes with cracks. Computers and Geotechnics. 77: 77-90.

[44] Zhu, D., 2000. The least upper-bound solutions for bearing capacity factor $N_\gamma$. Soils and foundations. 40(1): 123-129.

# Appendix-A

## A.1. *The expressions of the functions that constitute* $\dot{W}$

$$f_1(\theta_h, \theta_0) = \frac{1}{3(1+9\tan^2\phi')}\left\{(3\tan\phi'\cos\theta_h + \sin\theta_h)e^{3(\theta_h-\theta_0)\tan\phi'} - (3\tan\phi'\cos\theta_0 + \sin\theta_0)\right\}$$

$$f_2(\theta_h, \theta_0) = \frac{1}{6}\frac{L}{r_0}\left(2\cos\theta_0 - \frac{L}{r_0}\cos\omega\right)\sin(\theta_0 + \omega)$$

$$f_3(\theta_h, \theta_0) = \frac{1}{6}e^{(\theta_h-\theta_0)\tan\phi'}\left[\sin(\theta_h - \theta_0) - \frac{L}{r_0}\sin(\theta_h + \omega)\right]\left[\cos\theta_0 - \frac{L}{r_0}\cos\omega + \cos\theta_h e^{(\theta_h-\theta_0)\tan\phi'}\right]$$

$$f_4(\theta_h, \theta_0) = \frac{e^{3\tan\phi'(\theta_h-\theta_0)}(3\tan\phi'\sin\theta_h - \cos\theta_h) - 3\tan\phi'\sin\theta_0 + \cos\theta_0}{3(1+9\tan^2\phi')}$$

$$f_5(\theta_h, \theta_0) = \frac{1}{6}\frac{L}{r_0}\left(2\sin\theta_0 + \frac{L}{r_0}\sin\omega\right)\sin(\theta_0 + \omega)$$

$$f_6(\theta_h, \theta_0) = \frac{1}{6}e^{(\theta_h-\theta_0)\tan\phi'}\left[\sin(\theta_h - \theta_0) - \frac{L}{r_0}\sin(\theta_h + \omega)\right]\left[\sin\theta_0 + \frac{L}{r_0}\sin\omega + \sin\theta_h e^{(\theta_h-\theta_0)\tan\phi'}\right]$$

and $L/r_0$ is a function of $\theta_0$ and $\theta_h$, as mentioned in Eq. (7)b.

## A.2. *Components of the total Dissipation rate* ($\dot{D}$)

| Due to internal cohesion ($\dot{D}_{cohesion}$) | Due to suction ($\dot{D}_{suction}$) |
|---|---|
| $\dot{D}_{cohesion} = \int_{\theta_0}^{\theta_h} c'(V\cos\varphi')\frac{rd\theta}{\cos\varphi'}$ | $\dot{D}_{suction} = \int_{\theta_0}^{\theta_h} c'_{apparent}(V\cos\varphi')\frac{rd\theta}{\cos\varphi'}$ |
| $= \int_{\theta_0}^{\theta_h} c'(r\Omega)rd\theta = \int_{\theta_0}^{\theta_h} c'\Omega r^2 d\theta$ | $= \int_{\theta_0}^{\theta_h} (-\sigma^s\tan\varphi')(V\cos\varphi')\frac{rd\theta}{\cos\varphi'}$ |
| $= \int_{\theta_0}^{\theta_h} c'\Omega r_0^2 e^{2(\theta-\theta_0)\tan\varphi'}d\theta$ | $= \int_{\theta_0}^{\theta_h} (-\sigma^s\tan\varphi')\Omega r^2 d\theta$ |
| $= \frac{c'r_0^2\Omega}{2\tan\varphi'}\left[e^{2(\theta_h-\theta_0)\tan\varphi'} - 1\right]$ | $= \int_{\theta_0}^{\theta_h} (-\sigma^s\tan\varphi')\Omega r_0^2 e^{2(\theta-\theta_0)\tan\varphi'}d\theta$ |

Here, $V = r\Omega$

# Appendix-B: Determining Stability number from the upper bound expression

(i) Seismic Loadings:

$$\gamma r_0^3 \Omega (f_1 - f_2 - f_3) + \gamma r_0^3 \Omega k_v (f_1 - f_2 - f_3) + \gamma r_0^3 \Omega k_h (f_4 - f_5 - f_6) = \dot{D}$$

$$\Rightarrow \gamma r_0^3 \Omega (1 + k_v)(f_1 - f_2 - f_3) + \gamma r_0^3 \Omega k_h (f_4 - f_5 - f_6) = \frac{c' r_0^2 \Omega}{2 \tan \varphi'} \left[ e^{2(\theta_h - \theta_0) \tan \varphi'} - 1 \right] + \int_{\theta_0}^{\theta_h} (-\sigma^s \tan \varphi') \Omega r_0^2 e^{2(\theta - \theta_0) \tan \varphi'} d\theta$$

After eliminating $\Omega r_0^2$: $\Rightarrow \gamma H \left( \frac{r_0}{H} \right) \left[ (1 + k_v)(f_1 - f_2 - f_3) + k_h (f_4 - f_5 - f_6) \right] = \frac{c'}{2 \tan \varphi'} \left[ e^{2(\theta_h - \theta_0) \tan \varphi'} - 1 \right] + \int_{\theta_0}^{\theta_h} (-\sigma^s \tan \varphi') e^{2(\theta - \theta_0) \tan \varphi'} d\theta$

$$\Rightarrow \left. \frac{\gamma H}{c'} \right|_{ps} = \left( \frac{\frac{1}{2 \tan \varphi'} \left[ e^{2(\theta_h - \theta_0) \tan \varphi'} - 1 \right] + \int_{\theta_0}^{\theta_h} (-\sigma^s \tan \varphi') e^{2(\theta - \theta_0) \tan \varphi'} d\theta}{\left( \frac{r_0}{H} \right) \left[ (1 + k_v)(f_1 - f_2 - f_3) + k_h (f_4 - f_5 - f_6) \right]} \right)$$

(ii) Surcharge Loadings:

$$\gamma r_0^3 \Omega (f_1 - f_2 - f_3) + p_s \Omega r_0^2 \left( \frac{L}{r_0} \cos \theta_0 - 0.5 \frac{L^2}{r_0^2} \cos \omega \right) = \dot{D}$$

$$\Rightarrow \gamma r_0^3 \Omega (f_1 - f_2 - f_3) + p_s \Omega r_0^2 \left( \frac{L}{r_0} \cos \theta_0 - 0.5 \frac{L^2}{r_0^2} \cos \omega \right) = \frac{c' r_0^2 \Omega}{2 \tan \varphi'} \left[ e^{2(\theta_h - \theta_0) \tan \varphi'} - 1 \right] + \int_{\theta_0}^{\theta_h} (-\sigma^s \tan \varphi') \Omega r_0^2 e^{2(\theta - \theta_0) \tan \varphi'} d\theta$$

After eliminating $\Omega r_0^2 \Rightarrow \gamma H \left( \frac{r_0}{H} \right)(f_1 - f_2 - f_3) + p_s \left( \frac{L}{r_0} \cos \theta_0 - 0.5 \frac{L^2}{r_0^2} \cos \omega \right) = \frac{c'}{2 \tan \varphi'} \left[ e^{2(\theta_h - \theta_0) \tan \varphi'} - 1 \right] + \int_{\theta_0}^{\theta_h} (-\sigma^s \tan \varphi') e^{2(\theta - \theta_0) \tan \varphi'} d\theta$

$$\Rightarrow \left. \frac{\gamma H}{c'} \right|_{sur} = \left( \frac{\frac{1}{2 \tan \varphi'} \left[ e^{2(\theta_h - \theta_0) \tan \varphi'} - 1 \right] + \int_{\theta_0}^{\theta_h} \frac{1}{c'} (-\sigma^s \tan \varphi') e^{2(\theta - \theta_0) \tan \varphi'} d\theta - \frac{p_s}{c'} \left( \frac{L}{r_0} \cos \theta_0 - 0.5 \frac{L^2}{r_0^2} \cos \omega \right)}{\left( \frac{r_0}{H} \right)(f_1 - f_2 - f_3)} \right)$$